\documentclass[journal]{IEEEtran}
\usepackage{amssymb}
\usepackage{amsmath}
\usepackage{cite}
\usepackage{url}
\usepackage{xcolor}
\usepackage{cite,graphicx,amsmath,amssymb}
\usepackage{subfigure}
\usepackage{citesort}
\usepackage{fancyhdr}
\usepackage{mdwmath}
\usepackage{mdwtab}
\usepackage{caption}
\usepackage{amsthm}
\usepackage{algorithm}
\usepackage{algorithmic}
\newtheorem{defn}{Definition}
\newtheorem{remark}{Remark}
\newtheorem{theorem}{Theorem}

\newtheorem{lemma}{Lemma}

\newtheorem{corollary}{Corollary}

\newtheorem{proposition}{Proposition}
\newcommand{\bm}[1]{\mbox{\boldmath{$#1$}}}

\begin{document}

\title{Non-Orthogonal Multiple Access for Air-to-Ground Communication}

\author{

Xidong~Mu,~\IEEEmembership{Student Member,~IEEE,}
        Yuanwei~Liu,~\IEEEmembership{Senior Member,~IEEE,}
        Li~Guo,~\IEEEmembership{Member,~IEEE,}
        and Jiaru~Lin,~\IEEEmembership{Member,~IEEE}

\thanks{This paper was presented in part at the IEEE Global Communications Conference (GLOBECOM), Waikoloa, HI, USA, Dec 9-13, 2019.~\cite{Mu2019}}
\thanks{X. Mu, L. Guo, and J. Lin are with School of Artificial Intelligence and Key Laboratory of Universal Wireless Communications, Ministry of Education, Beijing University of Posts and Telecommunications, Beijing, China (email:\{muxidong, guoli, jrLin\}@bupt.edu.cn).}
\thanks{Y. Liu is with Queen Mary University of London, London,
UK (email:yuanwei.liu@qmul.ac.uk).}
}

\maketitle
\begin{abstract}
  This paper investigates ground-aerial uplink non-orthogonal multiple access (NOMA) cellular networks. A rotary-wing unmanned aerial vehicle (UAV) user and multiple ground users (GUEs) are served by ground base stations (GBSs) by utilizing the uplink NOMA protocol. The UAV is dispatched to upload specific information bits to each target GBSs. Specifically, our goal is to minimize the UAV mission completion time by jointly optimizing the UAV trajectory and UAV-GBS association order while taking into account the UAV's interference to non-associated GBSs. The formulated problem is a mixed integer non-convex problem and involves infinite variables. To tackle this problem, we efficiently check the feasibility of the formulated problem by utilizing graph theory and topology theory. Next, we prove that the optimal UAV trajectory needs to satisfy the \emph{fly-hover-fly} structure. With this insight, we first design an efficient solution with predefined hovering locations by leveraging graph theory techniques. Furthermore, we propose an iterative UAV trajectory design by applying successive convex approximation (SCA) technique, which is guaranteed to coverage to a locally optimal solution. We demonstrate that the two proposed designs exhibit polynomial time complexity. Finally, numerical results show that: 1) the SCA based design outperforms the fly-hover-fly based design; 2) the UAV mission completion time is significantly minimized with proposed NOMA schemes compared with the orthogonal multiple access (OMA) scheme; 3) the increase of GUEs' quality of service (QoS) requirements will increase the UAV mission completion time.
\end{abstract}

\section{Introduction}

In recent years, unmanned aerial vehicles (UAVs), or referred as drones, has drawn significant attention due to the characteristics of high maneuverability and low cost~\cite{Zeng2016Wireless}. Many promising applications of UAVs have emerged such as cargo delivery, real-time video streaming, disaster rescue, communication enhancement and recovery, etc~\cite{UAV,Mozaffari2019Tutorial}. Compared with terrestrial communication links, UAVs flying at a high attitude usually have a high probability to establish line-of-sight (LoS) links~\cite{Hourani2014}, which greatly boosts the investigations on UAV communications. One one hand, UAVs equipped with communication devices can be deployed as aerial base stations (BSs)~\cite{Mozaffari2016Unmanned,Lyu2017Placement,Mozaffari2017IoT,Wu2018,Zeng2018Trajectory,Cai2018,Zeng2019Energy,Mozaffari2019,Duan2019trade}. Compared with terrestrial BSs, the mobility of UAVs in three dimension (3D) space can be exploited to enhance the system performance such as coverage area, communication throughput, etc. On the other hand, one promising application in UAV communications is to integrate UAVs as aerial users into cellular networks~\cite{Zeng2019Cellular,Bergh2016,Azari2017,Zhang2019,Challita2019}. The UAV users are served by ground base stations (GBSs) in existing cellular networks, which improves the performance of UAV-ground communications.

Non-orthogonal multiple access (NOMA) has been regarded as a promising technology in fifth generation (5G) communications~\cite{Liu2017,Cai_survey}. In power domain NOMA communications, multiple users are served in the same time/frequency resource and multiplexed in power levels. Successive interference cancellation (SIC) technique is invoked at receivers for extra interference cancelation and signal decoding. Compared with orthogonal multiple access (OMA), NOMA can greatly improve the spectrum efficiency when users' channel conditions yield a large difference~\cite{Chen2017Optimization}. Due to the superior spectrum efficiency feature and the ability of supporting massive connectivity, NOMA technique in conventional terrestrial communication systems has been extensively investigated in many aspects such as power allocation designs~\cite{Yang2016General,Ali2016}, user fairness and grouping schemes~\cite{Ding2016,Choi2016}, physical layer security~\cite{Liu2017Enhancing}, etc, which motive us to exploit the potential benefits of applying NOMA technology into UAV communications.

\subsection{Prior Works}

\subsubsection{Studies on UAV Communications}
The existing literatures on UAV communications mainly focus on enhancing system performance by exploiting the new introduced degree of freedom-\emph{UAV mobilty}. The optimal deployment or trajectory design of UAVs were investigated with various problems. Mozaffari {\em et al.}~\cite{Mozaffari2016Unmanned} studied the aerial BSs optimal deployment problem in coexistence with D2D communications, where the user outage probability in terms of the UAV altitude and the density of D2D users was analyzed. A spiral-based algorithm was proposed by Lyu {\em et al.}~\cite{Lyu2017Placement} for multiple UAVs deployment with the aim of ground users are covered with the minimum number of UAVs. {With the goal of minimizing the total transmit power of IoT devices, Mozaffari {\em et al.}~\cite{Mozaffari2017IoT} proposed an efficient UAV deployment approach for the UAV-enabled IoT network.} To further exploit the mobility of UAVs, Wu {\em et al.}~\cite{Wu2018} investigated the trajectory design in a Multi-UAV BSs network. In order to maximize the average rate of users, the trajectory of different UAVs, user scheduling and UAVs transmit power are optimized. Furthermore, Zeng {\em et al.}~\cite{Zeng2018Trajectory} studied a UAV-enabled multicasting system, where the mission completion time was minimized by UAV trajectory design, subject to a common file was successfully received by ground nodes. Cai {\em et al.}~\cite{Cai2018} investigated the UAV secure communications, where two UAVs are applied for information transmission and jamming, respectively. A novel path discretization method was proposed in~\cite{Zeng2019Energy} to deal with energy consumption minimization problem with a rotary-wing UAV. Swarm of UAVs acting as a virtual antenna array was proposed in~\cite{Mozaffari2019} for distributing data to ground users and a two step algorithm was designed to minimize the total service time. A novel heterogenous cloud based multi-UAV system was studied by Duan {\em et al.}~\cite{Duan2019trade}, with the objective of striking a power-vs-delay tradeoff. In contrast to rich works in UAV-assisted cellular communication, cellular-enabled UAV communication has received attention of researchers very recently. Bergh{\em et al.}~\cite{Bergh2016} presented some measurement and simulation results when the UAV was connected with existing LTE networks. It demonstrated that aerial interferences severely degraded the system performance. Azari {\em et al.}~\cite{Azari2017} studied the relationship between aerial interferences with various system configurations where aerial users and ground users coexist. Zhang {\em et al.}~\cite{Zhang2019} minimized the mission completion time via trajectory optimization with the cellular-connected UAV always maintaining its connection with GBSs. Challita {\em et al.}~\cite{Challita2019} investigated multi-UAVs path planning by applying reinforcement learning method to minimize UAVs' interferences to ground networks.

\subsubsection{Studies on Conventional NOMA Systems}
With the advantages of superior spectrum efficiency and user fairness, NOMA has been widely studied in conventional communication systems. Yang {\em et al.}~\cite{Yang2016General} proposed a power allocation scheme while considering different users' quality of service (QoS) requirement in both downlink and uplink NOMA scenarios. A dynamic user clustering and power allocation design in NOMA systems was proposed by Ali {\em et al.}~\cite{Ali2016} to maximize the system sum-throughput. Ding {\em et al.}~\cite{Ding2016} investigated the impact of different user grouping on system sum rate in fixed power allocation NOMA and cognitive-radio-inspired NOMA communication. User fairness was considered by Choi~\cite{Choi2016} while deciding power allocation among users in the downlink NOMA scenario. To further investigate the application of NOMA technique, Liu {\em et al.}~\cite{Liu2017Enhancing} investigated the secrecy performance of NOMA communication in large-scale networks, where artificial noise was invoked to enhance physical layer security. {Zhang {\em et al.}~\cite{ZhangUplink} proposed a power control scheme for uplink NOMA communications, where the outage probability and achievable sum rate are analyzed. Tabassum {\em et al.}~\cite{Tabassum} further analyzed the performance of multi-cell uplink NOMA systems under different SIC assumptions.} The power allocation and secondary user scheduling were investigated by Xu {\em et al.}~\cite{Xu2018}, where a video transmission model was established in cognitive NOMA wireless networks. Liu {\em et al.}~\cite{Liu2016Cooperative} invoked simultaneous wireless information and power transfer (SWIPT) technique in cooperative NOMA, where the stronger users can perform energy harvesting and act as relays to enhance the performance of weaker users. In order to improve the performance of cell-edge users, Ali {\em et al.}~\cite{Ali2018} considered coordinated multi-point (CoMP) transmission in multi-cell NOMA networks.

\subsubsection{Studies on UAV-NOMA Systems}
Some potential research directions of UAV NOMA communications are presented in~\cite{Liu2019}. The UAV deployed as an aerial BS to provide connectivity to GUEs with NOMA was investigated by Sohail {\em et al.}~\cite{Sohail2018}, where the power allocation and the UAV attitude are optimized to achieve maximum sum-rate. More particularly, Nasir {\em et al.}~\cite{Nasir2019} proposed an efficient algorithm to solve max-min rate problem by optimizing the UAV attitude, antenna beamwidth and resource allocation. To investigate multiple antennas technique in UAV NOMA communications, Hou {\em et al.}~\cite{Hou2019} invoked the stochastic geometry approach to analyze the system performance under LoS and non-line-of sight (NLoS) scenarios in the UAV MIMO-NOMA system. {Liu {\em et al.}~\cite{LiuPlacement} proposed a UAV deployment and power allocation scheme to enhance the performance of the UAV-NOMA network. Nguyen {\em et al.}~\cite{Nguyen} maximized the sum rate through resource allocation and decoding order design in UAV-assisted NOMA wireless backhaul networks. To tackle the interferences caused by UAV users, a aerial-ground interference mitigation framework named as uplink cooperative NOMA was designed by Mei {\em et al.}~\cite{Mei2019} with the structure of backhaul links among GBSs.} Duan {\em et al.}~\cite{Duan} studied the resource allocation problem in the Multi-UAVs uplink NOMA IoT system, where UAVs are assumed to hover in the air. {Hou {\em et al.}~\cite{Hou20192} analyzed the performance of multi-UAV aided NOMA networks under user-centric and UAV-centric strategies.} Furthermore, Cui {\em et al.}~\cite{Cui2018Joint} studied the mobile UAV NOMA system, where ground users were served by the UAV through downlink NOMA and the UAV trajectory was optimized to maximize the minimum achievable rate of GUEs. Sun {\em et al.}~\cite{SunCyclical} proposed a cyclical NOMA UAV-enabled network, where the UAV serves different ground users through the NOMA protocol in a cyclical manner.

\subsection{Motivation and Contributions}

While the aforementioned research contributions have laid a solid foundation on UAV-aided communications and NOMA technique, the investigations on ground-aerial NOMA communications are still quite open. To the best of our knowledge, there has been no existing works investigating mobile UAV users with uplink NOMA transmission. Notice that the trajectory design for UAV users in the uplink NOMA system is quite different from existing UAV BSs trajectory designs \cite{Zeng2018Trajectory,Cai2018,Zeng2019Energy,Zhang2019,Cui2018Joint}: 1) though the interference of UAVs can be removed with the SIC technique at the associated GBS, the non-associated GBSs/GUEs still suffer from UAVs' interference; 2) the locations of UAVs determine not only the achievable rate with the associated GBS but also the interference to other non-associated GBSs/GUEs, which motivates our main study in this work.

In this article, we consider the ground-aerial uplink NOMA cellular networks which consists of a rotary-wing UAV and multiple GUEs/GBSs. Specifically, uplink NOMA protocol is invoked at the GBSs for serving the UAV and GUEs. The UAV flies from the predefined initial location to final location while delivering required information bits to target GBSs. Meanwhile, the QoS requirements of GUEs need to be satisfied during the flight time. The main contributions of this paper are as follows:
\begin{itemize}
  \item  We propose a novel uplink NOMA framework for cellular-enabled UAV communication, where the UAV and GUEs are served by GBSs with uplink NOMA protocol. By utilizing this framework, we formulate the UAV mission completion time minimization problem by jointly optimizing the UAV trajectory and UAV-GBS association order.
  \item We design an efficient method to check the feasibility of the formulated problem. By leveraging graph theory and topology theory, a graph is properly constructed with the concept of pathwise connected. We mathematically prove that the feasibility of formulated problem can be checked by examining the connectivity of the constructed graph.
  \item We propose a \emph{fly-hover-fly} based design to obtain an efficient solution with the aid of floyd algorithm and travel salesman problem (TSP). The obtained result is proven to be asymptotically optimal as the required uploading information bits increase.
  \item We propose an efficient iterative UAV trajectory design for giving UAV-GBS association order, where successive convex approximation (SCA) technique is invoked to find a locally optimal solution.
  \item Numerical results demonstrate that: {1) the proposed SCA based trajectory design converges fast and satisfies the proposed fly-hover-fly structure;} 2) the proposed NOMA transmission scheme achieves a significant reduction in terms of UAV mission completion time compared with the OMA scheme;  3) higher QoS requirements of GUEs leads to a longer UAV mission completion time.
\end{itemize}

\subsection{Organization and Notations}

The rest of the paper is organized as follows. Section II presents the system model for ground-aerial uplink NOMA cellular networks. In Section III, the UAV mission completion time minimization problem is formulated and an efficient method is proposed to check the feasibility of formulated problem. In Section IV, some useful properties of optimal solution are revealed and a fly-hover-fly based solution is developed with predefined hovering locations. In Section V, an efficient iterative UAV trajectory design is proposed to find a locally optimal solution. Section VI presents the numerical results to validate the effectiveness of our proposed designs. Finally, Section VII concludes the paper.

\emph{Notations:} Scalars are denoted by lower-case letters, vectors are denoted by bold-face lower-case letters. ${\mathbb{R}^{M \times 1}}$ denotes the space of $M$-dimensional real-valued vector. For a vector ${\bf{a}}$, ${{\mathbf{a}}^T}$ denotes its transpose and $\left\| {\mathbf{a}} \right\|$ denotes its Euclidean norm. $\left\| {{\mathbf{\dot q}}\left( t \right)} \right\|$ denotes the derivative with respect to $t$. $ \cup $ and $ \cap $ are the union and intersection operation, respectively.

\section{System Model}

\begin{figure}[t!]
  \begin{center}
        \includegraphics[width=3in]{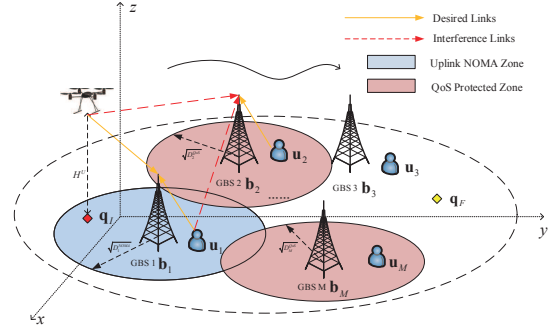}
        \caption{{Illustration of the ground-aerial uplink NOMA cellular networks.}}
        \label{System model}
  \end{center}
\end{figure}
As shown in Fig. \ref{System model}, the ground-aerial uplink NOMA cellular networks are considered, where a rotary-wing UAV has a mission of travelling from an initial location ${\bf{q}}_I$ to a final location ${\bf{q}}_F$. The UAV is dispatched to upload specific information bits to $M$ GBSs. It can be a practical UAV real-time streaming scenario. The maximum speed of the UAV is denoted as ${V_{\max }}$ and the UAV total mission completion time is denoted as $T$. {To invoke the NOMA transmission, full frequency reuse deployment is considered in the networks, where the whole bandwidth is available to every cell \cite{Tabassum,Ali2018}. For ease of presentation, we assume that each GBS serves one GUE. Our work can also be extended into the multiple GUEs scenario, which is described latter.} Denote the GBSs set as ${\cal M}_{BS}$ and the ground users set as ${\cal M}_{UE}$.

Without loss of generality, a 3D Cartesian coordinate system is considered. Assume that the UAV flies at a constant height of ${H^U}$ and the height of each GBSs is ${H^G}$. {In this paper, we assume that GUEs remain static with their locations know, such as IoT devices.} The coordinate of GBS $m$ is fixed at $\left( {x_m^G,y_m^G,{H^G}} \right)$ and its served GUE is located at $\left( {{x_m},{y_m},{H^{GUE}}} \right)$. Denote $\left( {{x}\left( t \right),{y}\left( t \right),{H^U}} \right)$, $0 \le t \le T$ as the trajectory of the UAV through the flight time. Then, the horizontal coordinates of the above locations are ${\bf{b}}_m = {\left[ {{x_m^G},{y_m^G}} \right]^T}$, ${\bf{u}}_m = {\left[ {{x_m},{y_m}} \right]^T}$, ${{\mathbf{q}}_I} = {\left[ {{x_I},{y_I}} \right]^T}$, ${{\mathbf{q}}_F} = {\left[ {{x_F},{y_F}} \right]^T}$ and ${{\bf{q}}}\left( t \right) = {\left[ {{x}\left( t \right),{y}\left( t \right)} \right]^T}$. The trajectory of the UAV needs to satisfy the following constraints,
\begin{align}\label{UAV Velocity Constraint}
&\left\| {{\mathbf{\dot q}}\left( t \right)} \right\| \le {V_{\max }},0 \le t \le T,\\
\label{UAV Initial Location Constraint}
&{{\bf{q}}}\left( 0 \right) = {\bf{q}}_I,{{\bf{q}}}\left( T \right) = {\bf{q}}_F.
\end{align}

For ease of exposition, the UAV and GUEs are assumed to be equipped with a single antenna. The antenna pattern of the GBSs is assumed to be horizontally omnidirectional but vertically directional and down-tilted to serve GUEs, the GBSs are assumed to receive UAV signals via the sidelobe~\cite{AzariCellular}. Let $g_m$ and $g_s$ denote as the the mainlobe and sidelobe gains of the GBS antennas. In the 3rd Generation Partnership Project (3GPP) technical report~\cite{3GPP_UAV}, a 100$\%$ LoS probability UAV-GBS channel can be achieved when the UAV's height is above a certain threshold in urban macro (UMa) and rural macro (RMa) scenarios. Therefore, we assume that the channel between the UAV and each GBS is dominated by the LoS link in this paper, since the UAV usually flies at a high altitude for safety consideration\footnote{{NLoS environment is also an important component for UAV communications, especially for the urban micro (UMi) scenario. Our future work will relax this LoS channel model assumption by considering the probabilistic LoS channel modelling such as in~\cite{Mozaffari2017IoT,Duan2019trade}.}}. It is also assumed that the Doppler effect caused by the UAV mobility is perfectly compensated at the receivers \cite{Synchronization}. For the air-to-ground channels, the pathloss between the UAV and GBS $m$ for a carrier frequency of $f$GHz in the UMa scenario is given by~\cite{3GPP_UAV}
\begin{align}\label{Channel gains between UAV and GBS m}
\begin{gathered}
  PL_m^{UAV} = 28 + 20\lg \left( f \right) \hfill \\
  \;\;\;\;\;\;\;\;\;\;\;\;\; + 22\lg \left( {\sqrt {{{\left( {{H^U} - {H^G}} \right)}^2} + {{\left\| {{\mathbf{q}}\left( t \right) - {{\mathbf{b}}_m}} \right\|}^2}} } \right). \hfill \\
\end{gathered}
\end{align}
The channel power gain between UAV and GBS $m$ can be expressed as
\begin{align}\label{Channel gains between UAV and GBS m}
{\left| {h_m^{UAV}\left( t \right)} \right|^2} = \frac{{{\rho _0}{g_s}}}{{{{\left( {{{\left( {{H^U} - {H^G}} \right)}^2} + {{\left\| {{\mathbf{q}}\left( t \right) - {{\mathbf{b}}_m}} \right\|}^2}} \right)}^{\frac{{\alpha}}{2}}}}},
\end{align}
where ${{\rho _0}}$ is the channel power gain at the reference distance of 1 meter and the path loss exponent $\alpha=2.2$.\\
\indent For the terrestrial channels, the pathloss between GUE $j$ and GBS $m$ for a carrier frequency of $f$GHz in the UMa scenario is given by~\cite{3GPP}
\begin{align}\label{Pathloss between user j and GBS m}
\begin{gathered}
  PL_{j,m}^{UE} = 32.4 + 20\lg \left( f \right) \hfill \\
  \;\;\;\;\;\;\;\;\;\;\;\;\; + 30\lg \left( {\sqrt {{{\left( {{H^G} - {H^{GUE}}} \right)}^2} + {{\left\| {{{\mathbf{u}}_j} - {{\mathbf{b}}_m}} \right\|}^2}} } \right). \hfill \\
\end{gathered}
\end{align}
In this paper, the channel power gain between GUE $j$ and GBS $m$ is approximated as ${\left| {h_{j,m}^{UE}} \right|^2} \approx {\mathbb{E}}\left[ {l_{j,m}^{UE}} \right]{g_i}{10^{ - \frac{{PL_{j,m}^{UE} + {\sigma _{SF}}}}{{10}}}}$, where ${{\sigma _{SF}}}$ denotes the shadow fading, $l_{j,m}^{UE}$ is the Rayleigh fading parameter and ${g_{j,m}} \in \left\{ {{g_s},{g_m}} \right\}$ represents the GBS antenna pattern gains\footnote{{In this paper, we assume that the GUEs are static, such as IoT devices. Therefore, the terrestrial channels might stay the same for a quite long time for the offline UAV trajectory design. Our future research would consider the online UAV trajectory design with mobile GUEs, which requires the prediction of the variety of terrestrial channels.}}. This approximation is based on the fact that the large scale fading dominates the channel gains, since the small-scale Rayleigh fading is on the different order of the magnitude compared to the distance-dependent path loss. A practical numerical example for comparison between the small-scale fading and path loss was provided in Chapter 2 of \cite{steele1999mobile}.

During the UAV mission completion time $T$, a binary variable ${a_{m}}\left( t \right) \in \left\{ {0,1} \right\}$ is defined to represent the UAV-GBS association state at time instant $t$. When the UAV is associated with GBS $m$ for data transmission at instant time $t$, ${a_{m}}\left( t \right) = 1$; otherwise, ${a_{m}}\left( t \right) = 0$. We assume that the UAV needs to maintain connectivity during $T$ and associate with at most one GBS at each time instant, we have $\sum\limits_{m = 1}^M {{a_m}\left( t \right)}  = 1,0 \le t \le T.$

Due to the strong UAV-GBS LoS links and the limited spectrum resource, uplink NOMA communication is considered. When the UAV and GUEs simultaneously transmit to associated GBSs in the same spectrum resource, the associated GBS first decodes the UAV's signal by treating its served GUE's signal as noise. After decoding the UAV's signal, the GBS decodes the GUE's signal with the UAV's signal subtracted with the SIC technology. Therefore, the received signal-to-interference-plus-noise (SINR) of the UAV at GBS $m$ at time instant $t$ can be expressed as
\begin{align}\label{One UAV SINR}
\gamma _m^{UAV}\left( t \right) = \frac{{{{\left| {h_m^{UAV}\left( t \right)} \right|}^2}{p^{UAV}}}}{{I_{\operatorname{intra} }^{UAV} + I_{\operatorname{inter} }^{UAV} + {\sigma ^2}}},
\end{align}
where $I_{\operatorname{intra} }^{UAV} = {\left| {h_{m,m}^{UE}} \right|^2}p_m^{UE}$ is the UAV suffered intra-cell interference, $I_{\operatorname{inter} }^{UAV} = \sum\limits_{j \ne m}^M {{{\left| {h_{j,m}^{UE}} \right|}^2}} p_j^{UE}$ is the UAV suffered inter-cell interference, ${p_{j}^{UE}}$ is the transmit power of GUE $j$, ${p^{UAV}}$ is the UAV transmit power and ${\sigma ^2}$ is the noise power.

The received SINR of GUE $m$ at GBS $m$ at time instant $t$ can be expressed as
\begin{align}\label{One UAV SINR user m}
\gamma _m^{UE}\left( t \right) = \frac{{{S_m}}}{{I_{\operatorname{inter} }^{UE} + {\sigma ^2}}},
\end{align}
where ${S_m} = {\left| {h_{m,m}^{UE}} \right|^2}p_m^{UE}$ is the signal of GUE $m$ and $I_{\operatorname{inter} }^{UE} = \left( {1 - {a_m}\left( t \right)} \right){\left| {h_m^{UAV}\left( t \right)} \right|^2}p_{}^{UAV} + \sum\limits_{j \ne m}^M {{{\left| {h_{j,m}^{UE}} \right|}^2}} p_j^{UE}$ is the GUE $m$ suffered inter-cell interference. {By integrating the UAV into the cellular network as an aerial user, the UAV interference is removed with the aid of the SIC technology at the associated GBS, e.g. ${a_{m}}\left( t \right) = 1$, and treated as inter-cell interference at the non-associated GBSs, e.g. ${a_{m}}\left( t \right) = 0$, which follows the conventional multi-cell uplink NOMA communications\footnote{The conventional multi-cell uplink NOMA with local SIC requires the minimum complexity to serve the UAV. Though the UAV interference to other non-associated GBSs is not canceled, it can be controlled by optimizing the UAV trajectory and UAV-GBS association orders which is known as the interference-aware UAV path planning [19].}~\cite{Tabassum}.}

Furthermore, when the UAV is associated with GBS $m$ at time instant $t$ for data transmission, the following constraint should be met to successfully perform the SIC at the GBS \cite{Yang2016General,Ali2016},
\begin{align}\label{Uplink NOMA}
{\left| {h_m^{UAV}\left( t \right)} \right|^2}{p^{UAV}} \ge {S_m},\;{\rm{if}}\;{a_m}\left( t \right) = 1.
\end{align}
Substituting \eqref{Channel gains between UAV and GBS m} in \eqref{Uplink NOMA}, constraint~\eqref{Uplink NOMA} can be further expressed as
\begin{align}\label{uplink NOMA zone}
0 \le {\left\| {{\mathbf{q}}\left( t \right) - {{\mathbf{b}}_m}} \right\|^2} \le D_m^{NOMA},
\end{align}
where $D_m^{NOMA} = {\left( {\frac{{{\beta _0}}}{{{S_m}}}} \right)^{\frac{2}{\alpha }}} - {H^2}$, ${\beta _0}={\rho _0}{g_s}p^{UAV}$ and $H = {H^U} - {H^G}$. \eqref{uplink NOMA zone} means if and only if the horizontal distance between the UAV and GBS $m$ is no larger than $\sqrt {D_m^{NOMA}} $, the UAV and GUE $m$ can be simultaneously served by GBS $m$ through uplink NOMA. We thus define a disk region on the horizontal plane centered at ${{\mathbf{b}}_m}$ with radius $\sqrt {D_m^{NOMA}} $ as the \emph{uplink NOMA zone}, {which is illustrated in Fig. \ref{System model}}. When the UAV is associated with GBS $m$ for data transmission, the uplink NOMA implementation requirement can be always satisfied if and only if its horizontal location lies in this region.

The instant achievable rate of the UAV at GBS $m$ is
\begin{align}\label{One UAV rate with GBS m}
{R_{m}^{UAV}}\left( t \right) = {{a_{m}}\left( t \right){{\log }_2}\left( {1 + {\gamma _{m}^{UAV}}\left( t \right)} \right)}.
\end{align}
The total uploaded information bits that the UAV transmits to GBS $m$ with a bandwidth $W$ during the mission completion time $T$ is expressed as
\begin{align}\label{One UAV throughput with GBS m}
\begin{array}{*{20}{l}}
  {{U_m} = \int\limits_0^T {WR_m^{UAV}\left( t \right)} dt} \\
  {\;\;\;\;\; = \int\limits_0^T {{a_m}\left( t \right)W{{\log }_2}\left( {1 + \frac{{{{\left| {h_m^{UAV}\left( t \right)} \right|}^2}{p^{UAV}}}}{{I_{\operatorname{intra} }^{UAV} + I_{\operatorname{inter} }^{UAV} + {\sigma ^2}}}} \right)} \;dt}.
\end{array}
\end{align}
Define ${\widetilde U_m}$ as the required information bits of GBS $m$ that the UAV needs to delivery, we have following constraints:
\begin{align}\label{One UAV throughput Constraint}
{U_m} \ge {\widetilde U_m},\;\;m \in {\cal M}_{BS}.
\end{align}
Similarly, the achievable rate of GUE $m$ at time instant $t$ is
\begin{align}\label{One UAV user m rate}
{R_{m}^{UE}}\left( t \right) =  {{{\log }_2}\left( {1 + {\gamma _{m}^{UE}}\left( t \right)} \right)}.
\end{align}
Define $\theta_m$ as the QoS requirement of GUE $m$. During the mission completion time $T$, the instant achievable rate constraint of each GUEs can be expressed as
\begin{align}\label{one USER QoS}
R_m^{UE}\left( t \right) \ge {\theta _m},0 \le t \le T,m \in {\mathcal{M}_{UE}}.
\end{align}
As described before, the instant achievable rate of GUEs depends on the UAV-GBS association state. We just need to concentrate on the interfering scenario, constraint~\eqref{one USER QoS} can be transformed into the following constraint,
\begin{align}\label{USER QoS 2}
{\log _2}\left( {1 + \frac{{{S_m}}}{{\frac{{{\beta _0}}}{{{{\left( {{{\left\| {{\mathbf{q}}\left( t \right) - {{\mathbf{b}}_m}} \right\|}^2} + {H^2}} \right)}^{\frac{\alpha }{2}}}}} + {I_m}}}} \right) \ge {\theta _m},\;{\rm{if}}\;{a_m}\left( t \right) = 0,
\end{align}
where ${I_m} = \sum\limits_{j \ne m}^M {{{\left| {h_{j,m}^{UE}} \right|}^2}} p_j^{UE} + {\sigma ^2}$. Constraint \eqref{USER QoS 2} can be further expressed as
\begin{align}\label{Protected zone}
{\left\| {{\mathbf{q}}\left( t \right) - {{\mathbf{b}}_m}} \right\|^2} \ge D_m^{QoS}.
\end{align}
where $D_m^{QoS} = {\left( {\frac{{{\beta _0}}}{{\frac{{{S_m}}}{{{2^{{\theta _m}}} - 1}} - {I_m}}}} \right)^{\frac{2}{\alpha}}} - {H^2}$. Similar with the definition of the uplink NOMA zone, \eqref{Protected zone} means when the UAV is not associated with GBS $m$, the horizontal distance between the UAV and GBS $m$ should not be smaller than $\sqrt {D_m^{QoS}} $ in order to guarantee the QoS requirement of GUE $m$. As illustrated in Fig. \ref{System model}, we define another disk region centered at ${{\mathbf{b}}_m}$ with radius $\sqrt {D_m^{QoS}} $ as the \emph{QoS protected zone} for GUE $m$ and the UAV cannot stay in when it is not associated with GBS $m$.
\begin{remark}\label{no user}
\emph{When the GBS is not serving any GUEs, the UAV can be served by the GBS directly. In this scenario, the radius of the uplink NOMA zone can be regarded to be a sufficiently large value and the radius of the QoS protected zone is 0.}
\end{remark}
\begin{remark}\label{multi user}
\emph{When the GBS serves multiple GUEs, the conventional uplink NOMA transmission scheme can be applied at the GBS \cite{Yang2016General}. In this scenario, the radius of the uplink NOMA zone is determined by the strongest GUE, while the radius of the QoS protected zone is determined by the most sensitive GUE.}
\end{remark}

Based on the concept of uplink NOMA zones and QoS protected zones, the feasible regions when the UAV is associated with GBS $m$ can be expressed as
\begin{align}\label{feasible region GBS m}
\begin{gathered}
  {\rm{if}}\;{a_m}\left( t \right) = 1,\;\;{\bf{q}}\left( t \right) \in {\mathcal{E}_m},\;m \in {{{\mathcal{M}}}_{BS}} \hfill \\
  {\rm{where}}\;\;\;\;{\mathcal{E}_m} = \left\{ {{\bf{q}} \in {\mathbb{R}^{2 \times 1}}:} \right.{\left\| {{\bf{q}} - {{\bf{b}}_m}} \right\|^2} \le D_m^{NOMA}, \hfill \\
  \left. {\;\;\;\;\;\;\;\;\;\;\;\;\;\;\;\;\;\;\;\;\;\;\;\;\;\;\;\left\| {{\bf{q}} - {{\bf{b}}_i}} \right\| \ge D_i^{QoS},i \ne m,i,m \in {{{\mathcal{M}}}_{BS}}} \right\}. \hfill \\
\end{gathered}
\end{align}
${\mathcal{E}_m}$ is a non-convex set in general and some properties of ${\mathcal{E}_m}$ in topology theory will be introduced in Section III.
\begin{remark}\label{multi SIC define}
\emph{It is worth noting that the SIC may also be performed at the non-associated GBSs as did in [39]. In our case, the condition for performing the SIC at the non-associated GBSs can be satisfied when the UAV stays at the overlap regions of different uplink NOMA zones. The feasible regions when the UAV is associated with GBS $m$ in this scheme can be expressed as ${{\mathcal{E}}}_m^{MS} = {{{\mathcal{E}}}_m} \cup \left( {\bigcup\limits_{i \ne m}^{{{\mathcal{M}}_{BS}}} {{{{\mathcal{E}}}_{m,i}}} } \right)$, where ${{{\mathcal{E}}}_{m,i}} = \left\{ {{\mathbf{q}} \in {\mathbb{R}^{2 \times 1}}:} \right.{\left\| {{\mathbf{q}} - {{\mathbf{b}}_m}} \right\|^2} \le D_m^{NOMA},\left. {{{\left\| {{\mathbf{q}} - {{\mathbf{b}}_i}} \right\|}^2} \le D_i^{NOMA}} \right\}$ denotes the overlap regions of two uplink NOMA zones. We defined this benchmark scheme as ``Multi-SIC'', which is also discussed in the following.}
\end{remark}

\section{Problem Formulation and Feasibility Check}

\subsection{Problem Formulation}

The purpose of this paper is to minimize the UAV mission complete time $T$ by jointly optimizing the trajectory of UAV ${\mathbf{Q}} = \left\{ {{\mathbf{q}}\left( t \right),0 \le t \le T} \right\}$ and the UAV-GBS association vectors ${\mathbf{A}} = \left\{ {{a_m}\left( t \right),0 \le t \le T,m \in {\mathcal{M}_{BS}}} \right\}$, while satisfying required upload information bits of each GBSs, each GUEs' QoS requirement, uplink NOMA implementation and the mobility of UAV. The optimization problem is formulated as
\begin{subequations}
\begin{align}\label{P1}
({\rm{P1}}):&\mathop {\min }\limits_{{\mathbf{Q}},{\mathbf{A},T}} \;\;T\\
\label{One UAV Initial Location Constraint}{\rm{s.t.}}\;\;&{{\bf{q}}}\left( 0 \right) = {\bf{q}}_I,{{\bf{q}}}\left( T \right) = {\bf{q}}_F,\left\| {{\mathbf{\dot q}}\left( t \right)} \right\| \le {V_{\max }},0 \le t \le T,\\
\label{UAV mission}&{U_m} \ge {\widetilde U_m},m \in {\cal M}_{BS},\\
\label{Uplink NOMA constraint}&\begin{gathered}
  {a_m}\left( t \right){\left\| {{\mathbf{q}}\left( t \right) - {{\mathbf{b}}_m}} \right\|^2} \le D_m^{NOMA},\; \hfill \\
  \;\;\;\;\;\;\;\;\;\;\;\;\;\;\;\;\;\;\;\forall m \in {{{\mathcal{M}}}_{BS}},0 \le t \le T, \hfill \\
\end{gathered} \\
\label{USER QoS}&\begin{gathered}
  {\left\| {{\mathbf{q}}\left( t \right) - {{\mathbf{b}}_m}} \right\|^2} \ge \left( {1 - {a_m}\left( t \right)} \right)D_m^{QoS}, \hfill \\
  \;\;\;\;\;\;\;\;\;\;\;\;\;\;\;\;\;\;\;\;\forall m \in {{{\mathcal{M}}}_{BS}},0 \le t \le T, \hfill \\
\end{gathered} \\
\label{One UAV Association Constraint}&\sum\limits_{m = 1}^M {{a_m}\left( t \right)}  = 1,0 \le t \le T,\\
\label{One UAV Association Constraint2}&{a_m}\left( t \right) \in \left\{ {0,1} \right\},\forall m \in {\mathcal{M}_{BS}}.
\end{align}
\end{subequations}
Constraints~\eqref{Uplink NOMA constraint} and~\eqref{USER QoS} represent the UAV is required to stay in the specific feasible regions when it is associated with different GBSs. There are two main reasons that make Problem (P1) challenging to solve. First, (P1) is a mixed integer non-convex problem due to the non-convex constraints~\eqref{UAV mission} and integer constraints~\eqref{One UAV Association Constraint2}. Constraints \eqref{Uplink NOMA constraint} and \eqref{USER QoS} further make ${\mathbf{Q}}$ and ${\mathbf{A}}$ coupled together. Second, the UAV trajectory ${\mathbf{Q}}$ and the UAV-GBS association vectors ${\mathbf{A}}$ are continuous functions of $t$, which make (P1) involve infinite number of optimization variables.

\subsection{Check the feasibility of (P1)}

Due to the existence of uplink NOMA zones and QoS protected zones, Problem (P1) may not be always feasible. For instance, when GUEs' QoS requirements are sufficiently high, the UAV can not be deployed owing to the introduced interference. Therefore, before solving Problem (P1), the feasibility of (P1) should be checked first. Recall the feasible regions of each GBSs defined in~\eqref{feasible region GBS m}, it is easy to prove that ${\mathcal{E}_m}$ is a topological space. Then, a definition called pathwise connected to describe the property of topological space is introduced.
\begin{defn}\label{definition of path-connectedness}
\emph{\cite{Topology}A topological space $\mathcal{E}$ is said to be pathwise connected if for any two points $x$ and $y$ in $\mathcal{E}$ there exists a continuous function $f$ from the unit interval $\left[ {0,1} \right]$ to $\mathcal{E}$ such that $f(0) = x$ and $f(1) = y$ (This function is called a path from $x$ to $y$).}
\end{defn}

Compared with the mathematic definition, it is much easier to determine whether ${\mathcal{E}_m}$ is pathwise connected in geometric view. In this paper, we first focus on the scenario that all ${\mathcal{E}_m}$ are pathwise connected. The solution to ${\mathcal{E}_m}$ which is not pathwise connected will be described latter. The infinite number of optimization variables make it difficult to check the feasibility of (P1). For any given UAV trajectory and association vectors $\left\{ {{\mathbf{q}}\left( t \right),{a_m}\left( t \right),m \in {\mathcal{M}_{BS}},0 \le t \le T} \right\}$, define
\begin{align}\label{association state}
\phi \left( t \right) \triangleq \left\{ {m \in {{{\mathcal{M}}}_{BS}}:{\mathbf{q}}\left( t \right) \in {\mathcal{E}_m},{a_m}\left( t \right) = 1} \right\}.
\end{align}
Constraints \eqref{Uplink NOMA constraint}-\eqref{One UAV Association Constraint2} are satisfied if and only if there exist $K$ critical time instances ${t_1},{t_2}, \cdots ,{t_K} $$\in \left( {0,T} \right)$, which denote the associated GBSs changes, i.e., $\phi\left( {{t_k} - \varepsilon } \right) \ne \phi\left( {{t_k}} \right)$ with any arbitrarily small $\varepsilon$. In other words, the GBS-UAV association order can be expressed as ${\bf{\Phi}}  = \left[ {{\phi_1},{\phi_2}, \cdots ,{\phi_{K + 1}}} \right]$ with ${\phi_k} \in {\mathcal{M}_{BS}}$. The UAV trajectory can be partitioned into $K+1$ portions and the UAV-GBS association remains unchanged during each portions. Denote ${t_0} = 0$ and ${t_{K+1}} = T$, for the $k$th portion, we have
\begin{align}\label{portion}
\phi\left( t \right) = {\phi_k},{\mathbf{q}}\left( t \right) \in {\mathcal{E}_{{\phi_k}}},{a_{{\phi_k}}}\left( t \right) = 1,{t_{k - 1}} \le t \le {t_k}.
\end{align}
Specifically, denote ${\mathbf{q}_{k,k + 1}} = \mathbf{q}\left( {{t_k}} \right)$ as the handover location from GBS $\phi_k$ to $\phi_{k+1}$. Then, we have the following proposition:
\begin{proposition}\label{feasible TO P1}
\emph{Problem (P1) is feasible if and only if there exists an UAV-GBS association order ${\bf{\Phi}}  = \left[ {{\phi_1},{\phi_2}, \cdots ,{\phi_{K}}} \right]$ that satisfies the following conditions:}
\begin{subequations}\label{feasible condition}
\begin{align}\label{Q1}
&{{\mathbf{q}}_I} \in {{{\mathcal{E}}}_{{\phi_1}}},{{\mathbf{q}}_F} \in {{{\mathcal{E}}}_{{\phi_{K}}}}\\
\label{QN_N+1}&{{{\mathcal{E}}}_{{\phi_k}}} \cap {{{\mathcal{E}}}_{{\phi_{k + 1}}}} \ne \emptyset ,k = 1,2, \cdots ,K-1\\
\label{A}&\left\{ {\bf{\Phi}} \right\} \supseteq {{{\mathcal{M}}}_{BS}},{\phi_k} \in {{{\mathcal{M}}}_{BS}},k = 1,2, \cdots ,K
\end{align}
\end{subequations}
\begin{proof}
See Appendix A.
\end{proof}
\end{proposition}

\begin{figure}[t!]
\centering
\subfigure[Horizontal locations of ${{\mathbf{q}}_I}$, ${{\mathbf{q}}_F}$ and $\left\{ {{{{\mathcal{E}}}_m}} \right\}$, $M=4$.]{\label{EX1}
\includegraphics[width= 2.5in, height=2.2in]{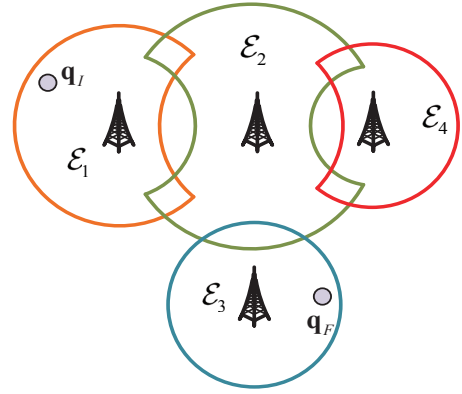}}
\subfigure[Constructed graph $G_0$.]{\label{EX2}
\includegraphics[width= 2.5in, height=1.8in]{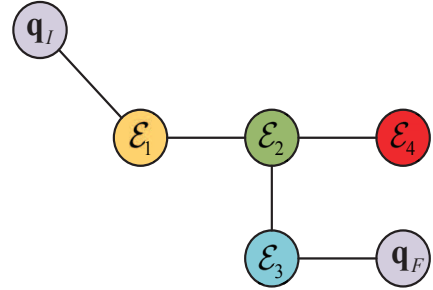}}
\setlength{\abovecaptionskip}{-0cm}
\caption{{Illustration of graph $G_0$ construction.}}\label{Graph0}
\end{figure}

With the definition of pathwise connected, \textbf{Proposition \ref{feasible TO P1}} implies Problem (P1) is feasible if and only if the topology space $\bigcup\limits_{m = 1}^M {{\mathcal{E}_m}} $ is pathwise connected and contains ${{\mathbf{q}}_I}$, ${{\mathbf{q}}_F}$. With this insight, we construct an undirected unweighted graph denoted by $G_0 = \left( {V_0,E_0} \right)$, where the vertices set $V_0$ is given by ${V_0} = \left\{ {{{\mathbf{q}}_I},{{\mathbf{q}}_F},{{{\mathcal{E}}}_1},{{{\mathcal{E}}}_2}, \cdots ,{{{\mathcal{E}}}_M}} \right\}.$ The edge set $E_0$ is defined as
\begin{align}\label{edge set}
\begin{gathered}
  {E_0} = \left\{ {\left( {{{\mathbf{q}}_I},{{{\mathcal{E}}}_m}} \right):\;{{\mathbf{q}}_I} \in {{{\mathcal{E}}}_m},m \in {{{\mathcal{M}}}_{BS}}} \right\} \hfill \\
  \;\;\;\;\;\;\; \cup \left\{ {\left( {{{\mathbf{q}}_F},{{{\mathcal{E}}}_m}} \right):\;{{\mathbf{q}}_F} \in {{{\mathcal{E}}}_m},m \in {{{\mathcal{M}}}_{BS}}} \right\} \hfill \\
  \;\;\;\;\;\;\; \cup \left\{ {\left( {{{{\mathcal{E}}}_i},{{{\mathcal{E}}}_j}} \right):{{{\mathcal{E}}}_i} \cap {{{\mathcal{E}}}_j} \ne \emptyset ,i \ne j \in {{{\mathcal{M}}}_{BS}}} \right\} \hfill. \\
\end{gathered}
\end{align}
For illustration, an example is provided in Fig. \ref{Graph0}. With the horizontal locations of all vertices as shown in Fig. \ref{EX1}, the corresponding graph $G_0$ is constructed, which is shown in Fig. \ref{EX2}. With graph {$G_0$}, the pathwise connectedness of $\bigcup\limits_{m = 1}^M {{\mathcal{E}_m}} $ is equivalent to the the connectivity of graph {$G_0$}. Compared with $\bigcup\limits_{m = 1}^M {{\mathcal{E}_m}} $, {$G_0$ has finite vertices which is composed of 2 points $\left\{ {{{\mathbf{q}}_I},{{\mathbf{q}}_F}} \right\}$ and $M$ point sets $\left\{ {{{{\mathcal{E}}}_1},{{{\mathcal{E}}}_2}, \cdots ,{{{\mathcal{E}}}_M}} \right\}$.} There exist many efficient algorithms to check graph connectivity, such as depth first search or breadth-first search \cite{graph}. As a result, the feasibility of Problem (P1) can be checked in a more tractable manner. For the ``Multi-SIC'' scheme, the proposed feasibility check method can be also applied with $\left\{ {{{\mathcal{E}}}_m^{MS}} \right\}$. In the following, we propose two efficient method to solve (P1) supposed that it has been verified to be feasible.

\section{Fly-Hover-Fly based Solution to Problem (P1)}

In this section, some useful insights on properties of the optimal solution to Problem (P1) are firstly revealed. With these insights, a fly-hover-fly based solution by utilizing graph theory techniques is proposed based on a properly designed graph. The properties and complexity analysis of the proposed algorithm are also provided.

\subsection{Properties of the Optimal Solution to (P1)}

Compared with fixed-wing UAVs, one of features of rotary-wing UAVs is the ability to hover in the air. Therefore, fly-hover-fly communication policy is an appealing solution in UAV trajectory design due to its simple structure. Base on the fly-hover-fly policy, we have the following theorem:
\begin{theorem}\label{fly-hover-fly policy}
\emph{Without lose of optimality to (P1), the optimal UAV trajectory can be assumed to be the following fly-hover-fly structure: Except hovering at specific locations, the UAV travels at the maximum speed $V_{max}$.}
\begin{proof}
See Appendix~B.
\end{proof}
\end{theorem}

It is worth noting that the proposed fly-hover-fly structure has a special case: when the hovering time is zero and the UAV always travels at $V_{max}$. Based on \textbf{Theorem \ref{fly-hover-fly policy}}, the total UAV mission completion time of (P1) can be expressed as
\begin{align}\label{fly-hover-fly time}
\begin{gathered}
  T\left( {{D_{fly}}} \right) = {T_{fly}} + {T_{hover}} \hfill \\
  \;\;\;\;\;\;\;\;\;\;\;\;\;\;\; = \sum\limits_{m = 1}^M {\left( {\frac{{{D_{fly,m}}}}{{{V_{\max }}}} + \frac{{{{\widetilde U}_m} - {U_{fly,m}}}}{{{R_{hover,m}}}}} \right)}  \hfill \\
  \;\;\;\;\;\;\;\;\;\;\;\;\;\;\; = \frac{{{D_{fly}}}}{{{V_{\max }}}} + \sum\limits_{m = 1}^M {\frac{{{{\widetilde U}_m} - {U_{fly,m}}}}{{{R_{hover,m}}}}}  \hfill \\
\end{gathered}
\end{align}
where ${T_{fly}}$ is the UAV total flying time, ${T_{hover}}$ is the UAV total hovering time, $D_{fly,m}$ is the total travelling distance when the UAV is associated with GBS $m$, $U_{fly,m}$ is the UAV uploaded information bits to GBS $m$ during travelling through $D_{fly,m}$. $R_{hover,m}$ is the communication rate when the UAV is associated with GBS $m$ and hovers at the corresponding hovering location. ${D_{fly}} = \sum\limits_{m = 1}^M {{D_{fly,m}}} $ is the total travelling distance through the mission. The monotonicity of $T\left( {{D_{fly}}} \right)$ is described with following proposition:
\begin{proposition}\label{monotonicity}
\emph{For any given ${{{\widetilde U}_m}}$ and ${{R_{hover,m}}}$, the mission completion time $T\left( {{D_{fly}}} \right)$ is a monotonically increasing function with respect to ${D_{fly}}$.}
\begin{proof}
\emph{We prove the Proposition~\ref{monotonicity} by showing for any $\Delta D > 0$, $T\left( {{D_{fly}} - \vartriangle D} \right) < T\left( {{D_{fly}}} \right)$ always holds. For any given $\Delta D > 0$,
\[T\left( {{D_{fly}} - \Delta D} \right) = \frac{{{D_{fly}} - \Delta D}}{{{V_{\max }}}} + \sum\limits_{m = 1}^M {\frac{{{{\widetilde U}_m} - {U_{fly,m}} + {U_{\Delta {D_m}}}}}{{{R_{hover,m}}}}}, \]
where ${{U_{\Delta D_m}}}$ represents the achieved throughput during $\frac{{\Delta D_m}}{{{V_{\max }}}}$. ${\Delta D_m}$ is the travelling distance when the UAV is associated with GBS $m$ through ${\Delta D}$, $\Delta D = \sum\limits_{m = 1}^M {\Delta {D_m}} $. With the definition of ${R_{hover,m}}$ in Appendix B, ${R_{hover,m}}$ is the highest communication rate when the UAV is associated with GBS $m$ along its trajectory. Thus, we have ${U_{\Delta {D_m}}} < \frac{{\Delta {D_m}}}{{{V_{\max }}}}{R_{hover,m}}$. Then,
\[\begin{gathered}
  T\left( {{D_{fly}} - \Delta D} \right) \hfill \\
   < \frac{{{D_{fly}} - \Delta D}}{{{V_{\max }}}} + \sum\limits_{m = 1}^M {\frac{{{{\widetilde U}_m} - {U_{fly,m}} + \frac{{\Delta {D_m}}}{{{V_{\max }}}}{R_{hover,m}}}}{{{R_{hover,m}}}}} \; \hfill \\
  \; = \frac{{{D_{fly}}}}{{{V_{\max }}}} + \sum\limits_{m = 1}^M {\frac{{{{\widetilde U}_m} - {U_{fly,m}}}}{{{R_{hover,m}}}}}  = T\left( {{D_{fly}}} \right). \hfill \\
\end{gathered} \]
The proof is completed.}
\end{proof}
\end{proposition}

\textbf{Proposition \ref{monotonicity}} implies with given hovering locations, ${T}\left( {{D_{fly}}} \right)$ achieves its minimum value when ${{D_{fly}}}$ is minimized. It means the optimal solution for Problem (P1) is to find the UAV-GBS association order which achieves shortest travelling distance from ${{\mathbf{q}}_I}$ to ${{\mathbf{q}}_F}$ visiting all optimal hovering locations. However, the optimal hovering locations are in general different with different ${{\widetilde U}_m}$ and it is non-trivial to determine. Assume that the $M$ optimal hovering locations to Problem (P1) are $\left\{ {{\mathbf{q}}_m^ * ,m \in {\mathcal{M}_{BS}}} \right\}$ and the optimal objective value of (P1) is $T\left( {\left\{ {{\mathbf{q}}_m^ * } \right\}} \right)$. In addition, define ${{\mathbf{q}}_m}$ as the location achieving the highest communication rate when UAV is associated with GBS $m$, such as ${\mathbf{q}_m} = \mathop {\max }\limits_{\mathbf{q} \in {\mathcal{E}_m}} R_m^{UAV}\left( \mathbf{q} \right)$. The objective value with $\left\{ {{{\mathbf{q}}_m}} \right\}$ is denoted as $T\left( {\left\{ {{\mathbf{q}}_m } \right\}} \right)$. Though $T\left( {\left\{ {{\mathbf{q}}_m } \right\}} \right)$ in general serves an upper bound for (P1) (i.e., $T\left( {\left\{ {{\mathbf{q}}_m^ * } \right\}} \right) \le T\left( {\left\{ {{\mathbf{q}}_m} \right\}} \right)$), it makes (P1) become more tractable to solve. Based on \textbf{Proposition \ref{monotonicity}}, the problem becomes equivalent to finding the shortest path from ${{\mathbf{q}}_I}$ to ${{\mathbf{q}}_F}$ while visiting all hovering locations $\left\{ {{{\mathbf{q}}_m}} \right\}$. In the following, an efficient algorithm is proposed with graph theory techniques to solve the above shortest path construction problem. Then, a high-quality solution is designed based on the obtained shortest path.

\subsection{Fly-Hover-Fly based Design with $\left\{ {{{\mathbf{q}}_m}} \right\}$}

\subsubsection{Finding UAV Hovering Locations ${{\mathbf{q}}_m}$ in each ${\mathcal{E}_m}$}
Since the UAV is assumed to fly at a constant height and the LoS channel model is considered. The hovering location ${{\mathbf{q}}_m}$ in ${\mathcal{E}_m}$ can be found by solving the following optimization problem:
\begin{subequations}
\begin{align}
\label{subproblem}({\rm{P2}}):&\mathop {\min }\limits_{\left\{ {{\mathbf{q}_m}} \right\}} \;\;\left\| {{\mathbf{q}_m} - {{\mathbf{b}}_m}} \right\|\\
\label{subproblem constraints1}{\rm{s.t.}}\;\;&{\left\| {{{\mathbf{q}}_m} - {{\mathbf{b}}_m}} \right\|^2} \le D_m^{NOMA},\\
\label{subproblem constraints2}&\left\| {{{\mathbf{q}}_m} - {{\mathbf{b}}_i}} \right\|^2 \ge D_i^{QoS},\forall i \in {\mathcal{M}_{BS}},i \ne m.
\end{align}
\end{subequations}
(P2) is non-convex due to the non-convex constraint~\eqref{subproblem constraints2}. In order to solve the above problem, we first reveal a useful property of the optimal hovering location for problem (P2) in the following proposition.
\begin{proposition}\label{optimal hovering position}
\emph{The optimal hovering location ${\mathbf{q}_m}$ for Problem (P2) has the following properties:\\
{\rm{(1)}}~If ${\mathbf{q}_m} = {{\mathbf{b}_m}}$ satisfies constraint~\eqref{subproblem constraints2}, the optimal hovering location is ${\mathbf{q}_m} = {{\mathbf{b}_m}}$;\\
{\rm{(2)}}~Otherwise, the optimal hovering location should satisfy the following condition: $\exists i \in {{{\mathcal{M}}}_{BS}},i \ne m,{\left\| {{{\mathbf{q}}_m} - {{\mathbf{b}}_i}} \right\|^2} = D_i^{QoS}$.}
\begin{proof}
See Appendix~C.
\end{proof}
\end{proposition}

By leveraging the above properties, Problem (P2) can be solved as following: First, check ${\mathbf{q}_m}= {{\mathbf{b}_m}}$ whether satisfying constraint~\eqref{subproblem constraints2}. If so, ${{\mathbf{b}_m}}$ is the optimal hovering location when the UAV is associated with GBS $m$. Otherwise, the optimal ${\mathbf{q}_m}$ can be obtained via one-dimensional search on the circle ${\left\| {{{\mathbf{q}}_m} - {{\mathbf{b}}_i}} \right\|^2} = D_i^{QoS},i \ne m \in {{{\mathcal{M}}}_{BS}}$ which satisfies constraint~\eqref{subproblem constraints1}. In this way, all $M$ hovering locations $\left\{ {{{\mathbf{q}}_m}} \right\},m \in {\mathcal{M}_{BS}}$ are obtained.

\subsubsection{Shortest Path Construction from ${{\mathbf{q}}_I}$ to ${{\mathbf{q}}_F}$ while Visiting All ${{\mathbf{q}}_m}$ with Graph Theory}
Different from the existing literatures~\cite{Zeng2018Trajectory,Zeng2019Energy}, the feasible regions ${\bigcup\limits_{m = 1}^M {{{{\mathcal{E}}}_m}} }$ is non-convex and the TSP cannot be applied straightly. First, we have the following theorem to describe the structure of desired shortest path.
\begin{theorem}\label{shortest path structure}
\emph{For the shortest path from ${{\mathbf{q}}_I}$ to ${{\mathbf{q}}_F}$ while visiting all ${{\mathbf{q}}_m}$, the path between any two neighbor points must be the shortest path between them.}
\begin{proof}
\emph{Theorem \ref{shortest path structure} can be shown by construction. Specifically, for any path with a given visiting order, if the path between any two neighbor points is not their shortest path. We can always construct a new path with the existing visiting order by replacing the previous path with the shortest path between two neighbor points and achieve shorter total path length. The proof is completed.}
\end{proof}
\end{theorem}

Based on $G_0$, we construct another undirected weighted graph denoted as $G_1 = \left( {V_1,E_1} \right)$  by replacing ${\mathcal{E}_m}$ with ${{\mathbf{q}}_m}$, where the vertices set $V_1$ is given by $V_1 = \left\{ {{{\mathbf{q}}_I},{{\mathbf{q}}_1},{{\mathbf{q}}_2}, \cdots ,{{\mathbf{q}}_M},{{\mathbf{q}}_F}} \right\}.$ The edge set $E_1$ is given by $E_1 = \left\{ {\left( {{{\mathbf{q}}_i},{{\mathbf{q}}_j}} \right),i \ne j \in \left\{ {\mathcal{M}}_{BS} \right\} \cup \left\{ {I,F} \right\}} \right\}.$ The weight of each edge is denoted as $d\left( {{{\mathbf{q}}_i},{{\mathbf{q}}_j}} \right)$, which represents the shortest path length between two vertices. We first focus on the method to calculate the shortest path length between two hovering locations. The shortest path length between ${{\mathbf{q}}_I}\left( {{{\mathbf{q}}_F}} \right)$ and $\left\{ {{{\mathbf{q}}_m}} \right\}$ can be treated as a special case. As graph $G_1$ is constructed based on graph $G_0$, the existence of edge $\left( {{{\mathbf{q}}_i},{{\mathbf{q}}_j}} \right)$ means ${\mathcal{E}_i} \cap {\mathcal{E}_j} \ne \emptyset $. Based on the concept of pathwise connected, we can always construct a feasible path $f_{ij}$ from ${{{\mathbf{q}}_i}}$ to ${{{\mathbf{q}}_j}}$ with a handover position ${{\mathbf{q}}_{ij}} \in {\mathcal{E}_i} \cap {\mathcal{E}_j}$. However, $f_{ij}$ is a continuous function and involves infinite number of variables, the shortest path length of $f_{ij}$ is difficult to calculate. To tackle this challenge, we invoke path discretization method to approximate the shortest path length between ${{{\mathbf{q}}_i}}$ and ${{{\mathbf{q}}_j}}$.

With path discretization method, the continuous path $f_{ij}$ is discretized into $2N_s$ line segments by $2N_s+1$ waypoints, where ${\mathbf{q}}\left[ 1 \right] = {{\mathbf{q}}_i}$, ${\mathbf{q}}\left[ 2N_s+1 \right] = {{\mathbf{q}}_j}$ and ${\mathbf{q}}\left[ N_s+1 \right]$ is the handover point. In order to achieve good approximation, we have following constraints:
\begin{align}\label{path discretization}
\left\| {{\mathbf{q}}\left[ {n + 1} \right] - {\mathbf{q}}\left[ n \right]} \right\| \le \delta ,n = 1, \cdots ,2N_s,
\end{align}
where $\delta $ is chosen sufficiently small. Moreover, $N_s$ should be chosen large enough such that ${N_s}\delta $ is larger than the upper bound of shortest path length. The corresponding association vector is also discretized into $2NM$ variables, denoted as $\left\{ {{a_{mn}},m \in {\mathcal{M}_{BS}}} \right\}_{n = 1}^{2N_s}$. ${a_{mn}} \in \left\{ {0,1} \right\}$ represents the UAV-GBS association state during the $n$th line segment $\left( {{\mathbf{q}}\left[ n \right],{\mathbf{q}}\left[ {n + 1} \right]} \right)$. Since $d\left( {{{\mathbf{q}}_i},{{\mathbf{q}}_j}} \right)$ is the shortest path length between ${{\mathbf{q}}_i}$ and ${{\mathbf{q}}_j}$ when the UAV hands over from GBS $i$ to GBS $j$ via ${{\mathbf{q}}_{ij}}$. We have $\left\{ {{a_{in}} = 1,n = 1, \cdots ,N_s} \right\}$ and $\left\{ {{a_{jn}} = 1,n = N_s + 1, \cdots ,2N_s} \right\}$. Then, the minimum path length problem between ${\mathbf{q}_i}$ and ${\mathbf{q}_j}$ can be expressed as
\begin{subequations}\label{shortest path problem}
\begin{align}
\label{shortest path}({\rm{P3.1}}):&\mathop {\min }\limits_{\left\{ {{\mathbf{q}}\left[ n \right]} \right\}} \sum\limits_{n = 1}^{2N_s} {\left\| {{\mathbf{q}}\left[ {n + 1} \right] - {\mathbf{q}}\left[ n \right]} \right\|}\\
\label{qi}{\rm{s.t.}}\;\;&{\mathbf{q}}\left[ 1 \right] = {{\mathbf{q}}_i},{\mathbf{q}}\left[ 2N_s+1 \right] = {{\mathbf{q}}_j},\\
\label{path discretization1}
&\left\| {{\mathbf{q}}\left[ {n + 1} \right] - {\mathbf{q}}\left[ n \right]} \right\| \le \delta ,n = 1, \cdots ,2N_s,\\
\label{GBS11}&\begin{gathered}
  {a_{mn}}{\left\| {{\mathbf{q}}\left[ n \right] - {{\mathbf{b}}_m}} \right\|^2} \le D_m^{NOMA},n = 1, \cdots ,2{N_s}, \hfill \\
  \;\;\;\;\;\;\;\;\;\;\;\;\;\;\;\;\;\;\;\;\;\;\;\forall m \in {{{\mathcal{M}}}_{BS}}, \hfill \\
\end{gathered} \\
\label{GBS12}
&\begin{gathered}
  {\left\| {{\mathbf{q}}\left[ n \right] - {{\mathbf{b}}_m}} \right\|^2} \ge\left( {1 - {a_{mn}}} \right)D_m^{QoS},n = 1, \cdots ,2{N_s}, \hfill \\
  \;\;\;\;\;\;\;\;\;\;\;\;\;\;\;\;\;\;\;\;\;\;\;\;\forall m \in {{{\mathcal{M}}}_{BS}}. \hfill \\
\end{gathered}
\end{align}
\end{subequations}
Problem (P3.1) is non-convex due to the non-convex constraints~\eqref{GBS12}. Note that the left hand side (LHS) of~\eqref{GBS12} is a convex function with respect to ${{\mathbf{q}}\left[ n \right]}$, the first-order Taylor expansion can be applied to have the lower bound of ${\left\| {{\mathbf{q}}\left[ n \right] - {{\mathbf{b}}_m}} \right\|^2}$ at the given local point ${{{\mathbf{q}}^l}\left[ n \right]}$,
\begin{align}\label{lower boound of distance}
\begin{gathered}
  {\left\| {{\mathbf{q}}\left[ n \right] - {{\mathbf{b}}_m}} \right\|^2} \ge {\left\| {{{\mathbf{q}}^l}\left[ n \right] - {{\mathbf{b}}_m}} \right\|^2} \hfill \\
  \;\;\;\;\;\;\;\;\;\;\;\;\;\;\;\;\;\; + 2{\left( {{{\mathbf{q}}^l}\left[ n \right] - {{\mathbf{b}}_i}} \right)^T} \times \left( {{\mathbf{q}}\left[ n \right] - {{\mathbf{q}}^l}\left[ n \right]} \right). \hfill \\
\end{gathered}
\end{align}
Then,~\eqref{GBS12} can be replaced with their low bounds at given points with \eqref{lower boound of distance}. (P3.1) can be written as
\begin{subequations}\label{shortest path problem2}
\begin{align}
\label{shortest path1}(&{\rm{P3.2}}):\mathop {\min }\limits_{\left\{ {{\mathbf{q}}\left[ n \right]} \right\}} \sum\limits_{n = 1}^{2N_s} {\left\| {{\mathbf{q}}\left[ {n + 1} \right] - {\mathbf{q}}\left[ n \right]} \right\|}\\
\label{approximated GBS121}{\rm{s.t.}}\;\;&
\begin{gathered}
  {\left\| {{{\mathbf{q}}^l}\left[ n \right] - {{\mathbf{b}}_m}} \right\|^2} + 2{\left( {{{\mathbf{q}}^l}\left[ n \right] - {{\mathbf{b}}_m}} \right)^T} \times \left( {{\mathbf{q}}\left[ n \right] - {{\mathbf{q}}^l}\left[ n \right]} \right) \hfill \\
  \;\;\;\;\;\;\; \ge \left( {1 - {a_{mn}}} \right)D_m^{QoS},n = 1, \cdots ,2N_s,\forall m \in {{{\mathcal{M}}}_{BS}} \hfill \\
\end{gathered} \\
\label{approximated GBS231}&\eqref{qi}-\eqref{GBS11}.
\end{align}
\end{subequations}
Problem (P3.2) is a convex problem which can be efficiently solved by using standard convex optimization software toolbox such as CVX~\cite{cvx}. Due to the lower bound in~\eqref{lower boound of distance}, the feasible region of (P3.2) is a subset of (P3.1). The optimal value of (P3.2) serves an upper bound to that of (P3.2). By updating the local points with the obtained results after each iteration, the optimal value of (P3.2) will converge to a locally optimal solution in a non-increasing manner. Thus, we have the approximate shortest path length $d\left( {{{\mathbf{q}}_i},{{\mathbf{q}}_j}} \right)$ and the corresponding path. Regarding the shortest path length between ${{\mathbf{q}}_I}\left( {{{\mathbf{q}}_F}} \right)$ and $\left\{ {{{\mathbf{q}}_m}} \right\}$, we can just remove the handover point and make $\left\{ {{a_{mn}} = 1,n = 1, \cdots ,2N_s} \right\}$ and solve (P3.2). Specifically, (P3.2) contains $2N_s$ second-order cone (SOC) constraints of size 4, $2N_s+1$ SOC constraints of size 2 and $2N_s+5$ linear inequality constraints of size 2.  The total number of optimization variables is $n_1 = 2N_s + 1$. The total complexity of solving (P3.2) with interior-point method is given by ${I_{iter,1}}{n_1}\sqrt {2\left( {14{N_s} + 7} \right)} \left( {\left( {4{n_1} + 8} \right)\left( {2{N_s} + 5} \right) + 40{N_s} + 4 + n_1^2} \right)$, i.e., $\mathcal{O}\left( {I_{iter,1}}{N_s^{3.5}} \right)$, where ${I_{iter,1}}$ denotes the number of iterations~\cite{WangComplexity}.

Next, we propose an approach to construct the shortest path from ${{\mathbf{q}}_I}$ to ${{\mathbf{q}}_F}$ while visiting all ${{\mathbf{q}}_m}$ based on graph $G_1$ by leveraging Floyd algorithm \cite{graph} and standard TSP. The shortest path construction problem can be regarded as a modified TSP problem. In the standard TSP~\cite{Zeng2018Trajectory}, the salesman needs to start and end with the same city and visit other cities only once. The purpose is to find the minimum traveling distance while visiting all the cities. Though the standard TSP is a NP-hard problem, there are many efficient algorithm to solve the standard TSP with time complexity $\mathcal{O}\left( {M}^2 \right)$~\cite{LaporteThe,Rego}. However, in our case, the salesman (UAV) is required to start and end with two different cities (vertices) and visit different cities (vertices) at least once. In order to solve our problem, we try to convert the modified TSP into a standard case, which can be efficiently solved.

Recall from \textbf{Theorem~\ref{shortest path structure}}, we first apply Floyd algorithm with time complexity $\mathcal{O}\left( {M}^3 \right)$ to $G_1$ to find the shortest path between any two different vertices and update $G_1$ with the new edge weight. We obtain a new graph denoted as $G_f$, whose each edge weight represents the shortest path between any two different vertices, and a path index matrix $P$, which contains the shortest path route between two different vertices. Based on $G_f$, a dummy point is added, whose distance to ${{\mathbf{q}}_I}$ and ${{\mathbf{q}}_F}$ is zero, and to other vertices is infinite large. Denote the graph with the dummy point as $G_d$. Our shortest path construction problem can be solved by treating $G_d$ as a standard TSP starting and ending with the dummy point. With the desired path after solving the standard TSP, we can reconstruct the origin path between two points with the path route in $P$ and remove the two edges connected with the dummy point. As a result, the shortest path $\left\{ {{\mathbf{q}}\left[ n \right]} \right\}_{n = 1}^{{N_{total}} + 1}$ and corresponding UAV-GBS association vectors $\left\{ {{a_{mn}},m \in {\mathcal{M}_{BS}}} \right\}_{n = 1}^{{N_{total}}}$ from ${{\mathbf{q}}_I}$ to ${{\mathbf{q}}_F}$ while visiting all ${{\mathbf{q}}_m}$ are obtained in path discretized form.

\subsubsection{Time Allocation based on the Proposed Fly-Hover-Fly Communication Structure}
With the shortest travelling path $\left\{ {{\mathbf{q}}\left[ n \right]} \right\}_{n = 1}^{{N_{total}} + 1}$ and corresponding UAV-GBS association vectors obtained, the fly-hover-fly based design remains to allocate time duration $\left\{ {{t_n}} \right\}_{n = 1}^{{N_{total}}}$ to each line segments and determine the hovering time at each ${\mathbf{q}}_m$. Based on \textbf{Theorem \ref{fly-hover-fly policy}}, the flying time of $n$th line segment is given by ${t_n} = \frac{{\left\| {{\mathbf{q}}\left[ {n + 1} \right] - {\mathbf{q}}\left[ n \right]} \right\|}}{{{V_{\max }}}}.$ Recall that $\delta $ is chosen sufficiently small such that the location of the UAV can be assumed remain unchanged during each line segments. Denote ${\eta _m} = \frac{{{\beta _0}}}{{{S_m} + {I_m}}}$, the uploaded information bits to GBS $m$ in $n$th line segment can be approximately calculated as ${U_{mn}} = {a_{mn}}{t_n}{\log _2}\left( {1 + \frac{{{\eta _m}}}{{{{\left( {{{\left\| {{\mathbf{q}}\left[ n \right] - {{\mathbf{b}}_m}} \right\|}^2} + {H^2}} \right)}^{\frac{\alpha }{2}}}}}} \right).$ Therefore, the hovering time at ${{\mathbf{q}}_m}$ is
\begin{align}\label{time minimization}
{t_{hover,m}} = \frac{{\max \left( {{{\widetilde U}_m} - \sum\limits_{n = 1}^{{N_{total}}} {{U_{mn}}} ,0} \right)}}{{{\log _2}\left( {1 + \frac{{{\eta _m}}}{{{{\left( {{{\left\| {{\mathbf{q}}\left[ n \right] - {{\mathbf{b}}_m}} \right\|}^2} + {H^2}} \right)}^{\frac{\alpha }{2}}}}}} \right)}}.
\end{align}
Above all, the corresponding UAV mission completion time with fly-hover-fly based design is $T = \sum\limits_{n = 1}^{{N_{total}}} {{t_n}}  + \sum\limits_{m = 1}^M {{t_{hover,m}}} $. The above algorithm is summarized in \textbf{Algorithm 1}. The computational complexity of \textbf{Algorithm 1} contains two parts. The first part is for solving (P3.2) with any two different vertices in $G_1$ with the complexity of $\mathcal{O}\left( {{I_{iter,1}}{M^2}N_s^{3.5}} \right)$. The second part is for solving shortest path construction problem with $G_1$ with the complexity of $\mathcal{O}\left( {M}^3 \right)$. Therefore, the total complexity of \textbf{Algorithm 1} is $\mathcal{O}\left( {{I_{iter,1}}{M^2}N_s^{3.5}} \right)$.

\begin{algorithm}[!h]\label{method1}
\caption{Fly-Hover-Fly based Solution}
 \hspace*{0.02in} \\
{\bf Input:} {$G_1$,$\left\{ {D_m^{NOMA},D_m^{QoS},{{\widetilde U}_m}} \right\},m \in {\mathcal{M}_{BS}}$.}\\
\vspace{-0.4cm}
\begin{algorithmic}[1]
\STATE Calculate the shortest path length $d\left( {{{\mathbf{q}}_i},{{\mathbf{q}}_j}} \right)$ between two different vertices by solving (P3.2) through iteration.
\STATE Apply floyd algorithm to $G_1$, $\left( {{G_f},P} \right) = {\rm{floyd}}\left( {{G_1}} \right)$.
\STATE Add dummy point to $G_f$ and solve the standard TSP with $G_d$.
\STATE Reconstruct the shortest path $\left\{ {{\mathbf{q}}\left[ n \right]} \right\}_{n = 1}^{{N_{total}} + 1}$ from ${{\mathbf{q}}_I}$ to ${{\mathbf{q}}_F}$ and obtain the UAV-GBS association vectors.
\STATE Time allocation based on \textbf{Theorem \ref{fly-hover-fly policy}} and calculate the UAV mission completion time.
\end{algorithmic}
{\bf Output:} the UAV mission completion time $T$, the UAV trajectory, UAV-GBS association vectors.
\end{algorithm}

\begin{remark}\label{remark:fly-hover-fly mode}
\emph{From \eqref{time minimization}, when ${\widetilde U_m} - \sum\limits_{n = 1}^{{N_{total}}} {{U_{mn}}}  < 0$, the obtained result is strictly suboptimal for (P1). When ${\widetilde U_m} \to \infty $, the fly-hover-fly based design with $\left\{ {{{\mathbf{q}}_m}} \right\}$ is asymptotically optimal. This is because compared with ${\widetilde U_m}$, the uploaded information bits while travelling {${U_{fly,m}}$} becomes negligible in this case. The minimum mission completion time is achieved with $M$ optimal hovering locations $\left\{ {{{\mathbf{q}}_m}} \right\}$ and the shortest travelling distance among them. }
\end{remark}

\section{SCA based Solution to Problem (P1)}

As mentioned before, the fly-hover-fly based design with $\left\{ {{{\mathbf{q}}_m}} \right\}$ in general serves an upper bound solution to (P1). Though the fly-hover-fly based design is asymptotically optimal for (P1) when ${\widetilde U_m} \to \infty $, the obtained result may not be tight enough especially when ${\widetilde U_m} - \sum\limits_{n = 1}^{{N_{total}}} {{U_{mn}}}  < 0$. To handle this problem, we propose an efficient algorithm to iteratively update the UAV trajectory by leveraging SCA technique. Specifically, we first transform Problem (P1) into discrete form with path discretization method. The UAV trajectory is dicreted into $N$ line segments with $N+1$ waypoints $\left\{ {{\mathbf{q}}\left[ n \right]} \right\}_{n = 1}^{N + 1}$ which satisfied constraint \eqref{path discretization}, where ${\mathbf{q}}\left[ 1 \right] = {{\mathbf{q}}_I}$ and ${\mathbf{q}}\left[ N+1 \right] = {{\mathbf{q}}_F}$. The time duration of $n$th line segments is $t_n$ and the UAV-GBS association vectors in $n$th line segments are $\left\{ {{a_{mn}}} \right\},{a_{mn}} \in \left\{ {0,1} \right\}$. Similarly, $N$ should be chosen sufficiently large such that $N\delta $ is larger than the upper bound of total required UAV travelling distance. The UAV mission completion minimization problem in path discretization form can be expressed as
\begin{subequations}\label{method 2}
\begin{align}
\label{method 2 ojective}({\rm{P4}}):&\mathop {\min }\limits_{\left\{ {{\mathbf{q}}\left[ n \right],{t_n},{a_{mn}}} \right\}} \;\;\sum\limits_{n = 1}^{{N}} {{t_n}}\\
\label{method 2 start}{\rm{s.t.}}\;\;&{\mathbf{q}}\left[ 1 \right] = {{\mathbf{q}}_I},{\mathbf{q}}\left[ {{N} + 1} \right] = {{\mathbf{q}}_F},\\
\label{path dis}&\left\| {{\mathbf{q}}\left[ {n + 1} \right] - {\mathbf{q}}\left[ n \right]} \right\| \le \min \left\{ {\delta ,{V_{\max }}{t_n}} \right\},n = 1, \cdots ,N,\\
\label{method 2 throughput}&\begin{gathered}
  \sum\limits_{n = 1}^N {{a_{mn}}{t_n}{{\log }_2}\left( {1 + \frac{{{\eta _m}}}{{{{\left\| {{\mathbf{q}}\left[ n \right] - {{\mathbf{b}}_m}} \right\|}^2} + {H^2}}}} \right)}  \ge {\widetilde U_m}, \hfill \\
  \;\;\;\;\;\;\;\;\;\;\;\;\;\;\;\;\;\;\;\;\;\;\;\;\;\;\;\;\;\;\;\;\;\;\;\;\;\;\;\;\;\;\;\forall m \in {{{\mathcal{M}}}_{BS}}, \hfill \\
\end{gathered} \\
\label{method 2 uplink NOMA}&\begin{gathered}
  {a_{mn}}{\left\| {{\mathbf{q}}\left[ n \right] - {{\mathbf{b}}_m}} \right\|^2} \le D_m^{NOMA}, \hfill \\
  \;\;\;\;\;\;\;\;\;\;\;\;\;\forall m \in {{{\mathcal{M}}}_{BS}},n = 1, \cdots ,N, \hfill \\
\end{gathered} \\
\label{method 2 QoS}&\begin{gathered}
  {\left\| {{\mathbf{q}}\left[ n \right] - {{\mathbf{b}}_m}} \right\|^2} \ge \left( {1 - {a_{mn}}} \right)D_m^{QoS}, \hfill \\
  \;\;\;\;\;\;\;\;\;\;\;\;\;\;\;\;\;\;\;\forall m \in {{{\mathcal{M}}}_{BS}},n = 1, \cdots ,N, \hfill \\
\end{gathered} \\
\label{method 2 constraint1}&\begin{gathered}
  \sum\limits_{m = 1}^M {{a_{mn}} = 1,{a_{mn}} \in \left\{ {0,1} \right\}} , \hfill \\
  \;\;\;\;\;\;\;\;\;\;\;\;\;\;\;\forall m \in {{{\mathcal{M}}}_{BS}},n = 1, \cdots ,N \hfill \\
\end{gathered} .
\end{align}
\end{subequations}
Though involving a finite number of variables, Problem (P4) is still a mixed integer non-convex problem which is difficult to solve. In the following, we concentrate on the UAV trajectory design under given UAV-GBS associated vectors. With any given UAV-GBS association vectors, the UAV trajectory design problem can be expressed as
\begin{subequations}\label{P2.1}
\begin{align}
\label{P2.1 ojective}({\rm{P5.1}}):&\mathop {\min }\limits_{\left\{ {{\mathbf{q}}\left[ n \right],{t_n}} \right\}} \;\;\sum\limits_{n = 1}^{{N}} {{t_n}}\\
\label{P2.1 constraints}{\rm{s.t.}}\;\;&\eqref{method 2 start}-\eqref{method 2 QoS}.
\end{align}
\end{subequations}
Problem (P5.1) is still non-convex due to the non-convex constraint \eqref{method 2 throughput} and \eqref{method 2 QoS}. To deal with the non-convex constraint \eqref{method 2 throughput}, we introduce slack variables $\left\{ {{\pi _{mn}},m \in {{{\mathcal{M}}}_{BS}},n = 1, \cdots ,N} \right\}$ and we have ${\pi _{mn}} = {\log _2}\left( {1 + \frac{{{\eta _m}}}{{{{\left( {{{\left\| {{\mathbf{q}}\left[ n \right] - {{\mathbf{b}}_m}} \right\|}^2} + {H^2}} \right)}^{\frac{\alpha }{2}}}}}} \right).
$ Then, Problem (P5.1) can be written as
\begin{subequations}\label{method 2.1}
\begin{align}
\label{method 2.1 ojective}({\rm{P5.2}}):&\mathop {\min }\limits_{\left\{ {{\mathbf{q}}\left[ n \right],{t_n},{\pi _{mn}}} \right\}} \;\;\sum\limits_{n = 1}^{{N}} {{t_n}}\\
\label{method 2.1 throughput}{\rm{s.t.}}\;\;&\sum\limits_{n = 1}^N {{a_{mn}}{t_n}{\pi _{mn}}}  \ge {\widetilde U_m},\forall m \in {{{\mathcal{M}}}_{BS}},\\
\label{method 2.1 Smn}&\begin{gathered}
  {\pi _{mn}} \le {\log _2}\left( {1 + \frac{{{\eta _m}}}{{{{\left\| {{\mathbf{q}}\left[ n \right] - {{\mathbf{b}}_m}} \right\|}^2} + {H^2}}}} \right), \hfill \\
  \;\;\;\;\;\;\;\;\;\;\;\;\;\;\;\;\;\;\;\forall m \in {{{\mathcal{M}}}_{BS}},n = 1, \cdots ,N + 1, \hfill \\
\end{gathered} \\
\label{method 2.1 constraint}&\eqref{method 2 start}-\eqref{path dis},\eqref{method 2 uplink NOMA},\eqref{method 2 QoS}.
\end{align}
\end{subequations}
It can be verified that (P5.2) achieves its optimal result when the constraint \eqref{method 2.1 Smn} is satisfied with equality. Otherwise, if any of constraint in \eqref{method 2.1 Smn} is satisfied with strictly inequality, we can always increase ${\pi _{mn}}$ to strictly equality and reduce the mission completion time. Thus, (P5.2) is equivalent to (P5.1). Problem (P5.2) is still non-convex due to non-convex constraints \eqref{method 2 QoS}, \eqref{method 2.1 throughput} and \eqref{method 2.1 Smn}. Fortunately, the three constraints can be tackled with SCA technique. In the former subsection, we have introduced how to deal with non-convex constraints like \eqref{method 2 QoS} with its lower bound at given points as \eqref{lower boound of distance}.

To tackle the non-convex constraint \eqref{method 2.1 throughput}, the LHS can be expressed as
\begin{align}\label{tn_pi}
{t_n}{\pi _{mn}} = \frac{{{{\left( {{t_n} + {\pi _{mn}}} \right)}^2}}}{2} - \frac{{t_n^2}}{2} - \frac{{\pi _{mn}^2}}{2}.
\end{align}
\noindent For the right hand side (RHS) of \eqref{tn_pi}, the first term is jointly convex with respect to ${{t_n}}$ and ${{\pi _{mn}}}$. Similarly, by applying the first-order Taylor expansion, the lower bound of RHS at given points $\left\{ {t_n^l,\pi _{mn}^l} \right\}$ is expressed as
\begin{align}\label{Smn lower bound}
\begin{gathered}
  \frac{{{{\left( {{t_n} + {\pi _{mn}}} \right)}^2}}}{2} \ge \lambda _{mn}^{lb} = \frac{{{{\left( {t_n^l + \pi _{mn}^l} \right)}^2}}}{2} \hfill \\
  \;\;\;\;\;\;\;\;\;\;\; + \left( {t_n^l + \pi _{mn}^l} \right)\left( {{t_n} - t_n^l} \right) + \left( {t_n^l + \pi _{mn}^l} \right)\left( {{\pi _{mn}} - \pi _{mn}^l} \right). \hfill \\
\end{gathered}
\end{align}
Moreover, to deal with the non-convex constraint \eqref{method 2.1 Smn}, the RHS is not concave with respect to ${{\mathbf{q}}\left[ n \right]}$, but it is a convex function with respect to ${{{\left\| {{\mathbf{q}}\left[ n \right] - {{\mathbf{b}}_m}} \right\|}^2}}$. Denote the lower bound as $R_{mn}^{lb}$ and we have
\begin{align}\label{Rmn lower bound}
\begin{gathered}
  {\log _2}\left( {1 + \frac{{{\eta _m}}}{{{{\left\| {{\mathbf{q}}\left[ n \right] - {{\mathbf{b}}_m}} \right\|}^2} + {H^2}}}} \right) \ge R_{mn}^{lb} \hfill \\
  \;\;\;\;\;\;\;\; = B_m^l\left[ n \right] - C_m^l\left[ n \right]\left( {{{\left\| {{\mathbf{q}}\left[ n \right] - {{\mathbf{b}}_m}} \right\|}^2} - {{\left\| {{{\mathbf{q}}^l}\left[ n \right] - {{\mathbf{b}}_m}} \right\|}^2}} \right), \hfill \\
\end{gathered}
\end{align}
where ${B_m^l}{\left[ n \right]} = {\log _2}\left( {1 + \frac{{{\eta _m}}}{{{{\left( {{{\left\| {{\mathbf{q}}^l\left[ n \right] - {{\mathbf{b}}_m}} \right\|}^2} + {H^2}} \right)}^{\frac{\alpha }{2}}}}}} \right)$ and $C_m^l\left[ n \right] = \frac{{\frac{\alpha }{2}\left( {{{\log }_2}e} \right){\eta _m}}}{{\left( {{{\left\| {{{\mathbf{q}}^l}\left[ n \right] - {{\mathbf{b}}_m}} \right\|}^2} + {H^2}} \right)\left[ {{{\left( {{{\left\| {{{\mathbf{q}}^l}\left[ n \right] - {{\mathbf{b}}_m}} \right\|}^2} + {H^2}} \right)}^{\frac{\alpha }{2}}} + {\eta _m}} \right]}}$. With \eqref{lower boound of distance}, \eqref{Smn lower bound} and \eqref{Rmn lower bound}, the non-convex constraints of Problem (P5.2) are replaced by their corresponding lower bounds with given points and (P5.2) is further expressed as following:
\begin{subequations}\label{method 2.2}
\begin{align}
\label{method 2.2 ojective}({\rm{P5.3}}):&\mathop {\min }\limits_{\left\{ {{\mathbf{q}}\left[ n \right],{t_n},{\pi _{mn}}} \right\}} \;\;\sum\limits_{n = 1}^{{N}} {{t_n}}\\
\label{method 2.2 mission}{\rm{s.t.}}\;\;&\sum\limits_{n = 1}^N {{a_{mn}}\left( {\lambda _{mn}^{\operatorname{lb} } - \frac{{t_n^2}}{2} - \frac{{\pi _{mn}^2}}{2}} \right)}  \ge {\widetilde U_m},\forall m \in {{{\mathcal{M}}}_{BS}},\\
\label{method 2.2 Smn}&{\pi _{mn}} \le R_{mn}^{lb},\forall m \in {{{\mathcal{M}}}_{BS}},n = 1, \cdots ,N,\\
\label{method 2.2 constraint1}&\begin{gathered}
  {\left\| {{{\mathbf{q}}^l}\left[ n \right] - {{\mathbf{b}}_m}} \right\|^2} + 2{\left( {{{\mathbf{q}}^l}\left[ n \right] - {{\mathbf{b}}_m}} \right)^T} \times \left( {{\mathbf{q}}\left[ n \right] - {{\mathbf{q}}^l}\left[ n \right]} \right) \hfill \\
  \;\;\;\;\;\; \ge {\left( 1-a_{mn} \right)}D_m^{QoS},\forall m \in {{{\mathcal{M}}}_{BS}},n = 1, \cdots ,N, \hfill \\
\end{gathered}\\
\label{method 2.2 constraint2}&\eqref{method 2 start}-\eqref{path dis},\eqref{method 2 uplink NOMA}.
\end{align}
\end{subequations}
It is easy to verify (P5.3) is a convex problem which can be efficiently solved by using standard convex optimization software toolbox such as CVX~\cite{cvx}. It is worth noting that the optimal value of (P5.3) in general serves an upper bound for that of (P5.1) due to invoking the lower bounds in \eqref{lower boound of distance}, \eqref{Smn lower bound} and \eqref{Rmn lower bound}. It makes any feasible solution to (P5.3) is also feasible for (P5.1), but the reverse not necessarily hold. It suggests that the feasible set of (P5.3) is a subset of (P5.1). After each iteration, the local points are updated with the obtained results and the optimal result of (P5.3) will converge to a locally optimal solution in a non-increasing manner. The proposed SCA based trajectory design is summarized in \textbf{Algorithm 2}.

\textbf{Algorithm 2} indicates that Problem (P5.1) can be efficiently solved through iteration with given UAV-GBS association order and an initial feasible UAV trajectory. Therefore, one straightforward solution to Problem (P4) is exhaustively finding all feasible UAV-GBS association order, and choosing the minimum result after solving (P5.1). However, this method is hard to implement. On one hand, finding all possible UAV-GBS association order requires a prohibitive complexity, e.g., $\mathcal{O}\left( {M!} \right)$, which is unacceptable for moderate values of $M$. On the other hand, solving problem (P5.1) needs an initial feasible solution to start the iteration. The coverage speed and obtained objective value in general depend on the initialization. To achieve high-quality performance and fast coverage speed, we choose the fly-hover-fly based solution in Section IV for initialization. Specifically, Problem (P5.3) involves $N$ SOC constraints of size 4, $N$ SOC constraints of size 5, $MN$ SOC constraints of size 2, $M$ SOC constraints of size $2N$, $MN$ SOC constraints of size 3 and $MN+4$ linear inequality constraints (2 linear equality constraint) of size 2. The total number of optimization variables is $n_2=MN+3N+2$. The total complexity of solving (P5.3) with interior-point method is given by ${I_{iter,1}}{n_2}\sqrt {2\left( {8MN + 9N + 4} \right)} \left( {8\left( {MN + 4} \right) + 4{n_2}\left( {MN + 4} \right)  } \right.$\\$\left. +{4M{N^2} + 13MN + 41N + n_2^2} \right)$, i.e., $\mathcal{O}\left( {{I_{iter,2}}{M^{3.5}}{N^{3.5}}} \right)$, where ${I_{iter,2}}$ denote the number of iterations{~\cite{WangComplexity}}.
\begin{algorithm}[!h]\label{method1}
\caption{SCA based trajectory design}
 \hspace*{0.02in} \\
{\bf Input:} {Feasible UAV-GBS association vectors $\left\{ {{a_{mn}}} \right\}$ and the UAV trajectory $\left\{ {{\mathbf{q}}\left[ n \right]} \right\}$, $\left\{ {{t_n}} \right\}$ to (P5.1).}\\
\vspace{-0.4cm}
\begin{algorithmic}[1]
\STATE {\bf Initialization}: {$l=0$, ${{\mathbf{q}}^l}\left[ n \right] = {\mathbf{q}}\left[ n \right]$, $t_n^l = {t_n}$}.
\STATE {\bf repeat}
\STATE Calculate ${\pi _{mn}^l}$.
\STATE Obtain the optimal solution ${{\mathbf{q}}^o}\left[ n \right]$, $t_n^o$ by solving (P5.3).
\STATE $l=l+1$, ${{\mathbf{q}}^l}\left[ n \right] = {{\mathbf{q}}^o}\left[ n \right]$, $t_n^l = t_n^o$.
\STATE {\bf until} the fractional decrease of the objective value is below a threshold $\varepsilon  > 0$.
\end{algorithmic}
{\bf Output:} The UAV trajectory and the mission completion time $\sum\limits_{n = 1}^{{N}} {{t_n}}$.
\end{algorithm}
\begin{remark}\label{remark:SCA mode}
\emph{We choose the fly-hover-fly based solution as initial points for the SCA based trajectory design to solve Problem (P5.1). It suggests that the obtained result of the SCA based trajectory design will be no larger than the fly-hover-fly based design. Meanwhile, the asymptotically optimal feature of the fly-hover-fly based design ensures the high-quality performance \footnote{{It usually requires a large amount of information bits transmission in practical UAV applications, especially for the UAV mission-specific payload communication \cite{Zhang2019}. Other initialization schemes may enhance the performance of the SCA based scheme when the required information bit is low, which is set aside for our future work.}} and fast coverage speed of the SCA based trajectory design, which are validated in numerical results.}
\end{remark}
\begin{remark}\label{remark:multi-SIC}
\emph{The condition that $D_m^{NOMA} > D_m^{QoS},\forall m \in {{\mathcal{M}}_{BS}}$ should be always satisfied as long as Problem (P1) is feasible. Otherwise, it is impossible to construct a pathwise connected topology space $\bigcup\limits_{m = 1}^M {{\mathcal{E}_m}}$ (or $\bigcup\limits_{m = 1}^M {{{\mathcal{E}}}_m^{MS}} $). With this condition, it is evident that the feasible region of each GBS in the ``Multi-SIC'' scheme ${{{\mathcal{E}}}_m^{MS}}$ is the corresponding uplink NOMA zone (i.e., $\left\{ {{\mathbf{q}} \in {\mathbb{R}^{2 \times 1}}:{{\left\| {{\mathbf{q}} - {{\mathbf{b}}_m}} \right\|}^2} \le D_m^{NOMA}} \right\}$). Therefore, our proposed algorithms can be applied for the ``Multi-SIC'' scheme by only considering the uplink NOMA zones. The performance of the ``Multi-SIC'' scheme will be provided in Section VI.}
\end{remark}
\begin{remark}\label{not pathwise}
\emph{Although our proposed algorithm is based on the condition that each ${\cal E}_m$ is pathwise connected, it can be also applied when ${\cal E}_m$ is not pathwise connected. For instance, assume that ${\cal E}_m, {m \in {{{\mathcal{M}}}_{BS}}}$ is not pathwise connected, we can apply our proposed algorithms by replacing ${\cal E}_m$ with a pathwise connected topological space ${\widehat {\mathcal{E}}_m}$, which satisfies ${\widehat {\mathcal{E}}_m} \subseteq {{\mathcal{E}}_m}$. However, the obtained objective value in this scenario serves an upper bound due to the approximation.}
\end{remark}

\section{Numerical Examples}

\begin{figure*}[t!]
\centering
\subfigure[$\theta$=0.3 bit/s/Hz and $U$=20 Mbits.]{\label{QoS03Data20}
\includegraphics[width= 2.3in, height=1.8in]{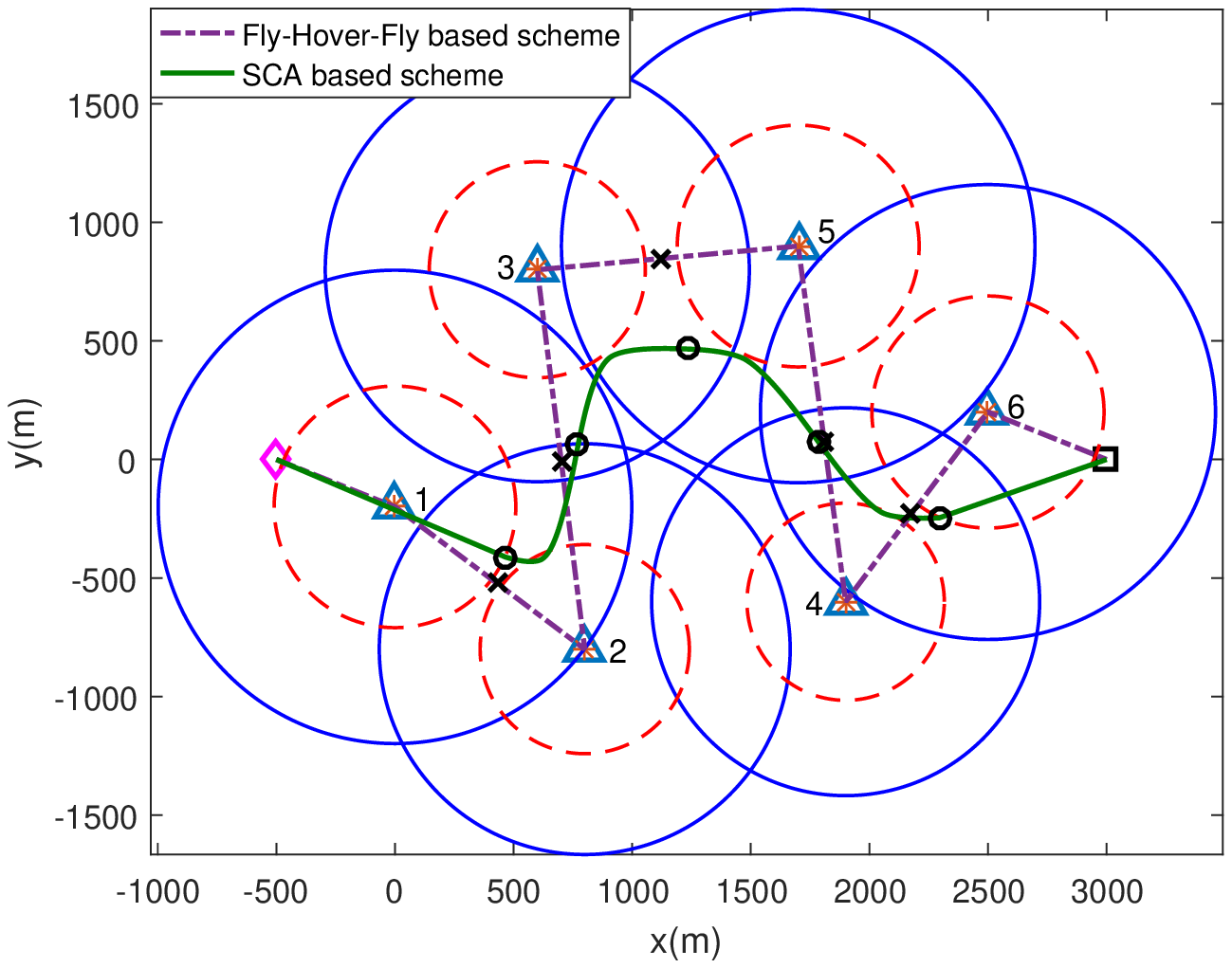}}
\subfigure[$\theta$=0.8 bit/s/Hz and $U$=20 Mbits.]{\label{QoS08Data20}
\includegraphics[width= 2.3in, height=1.8in]{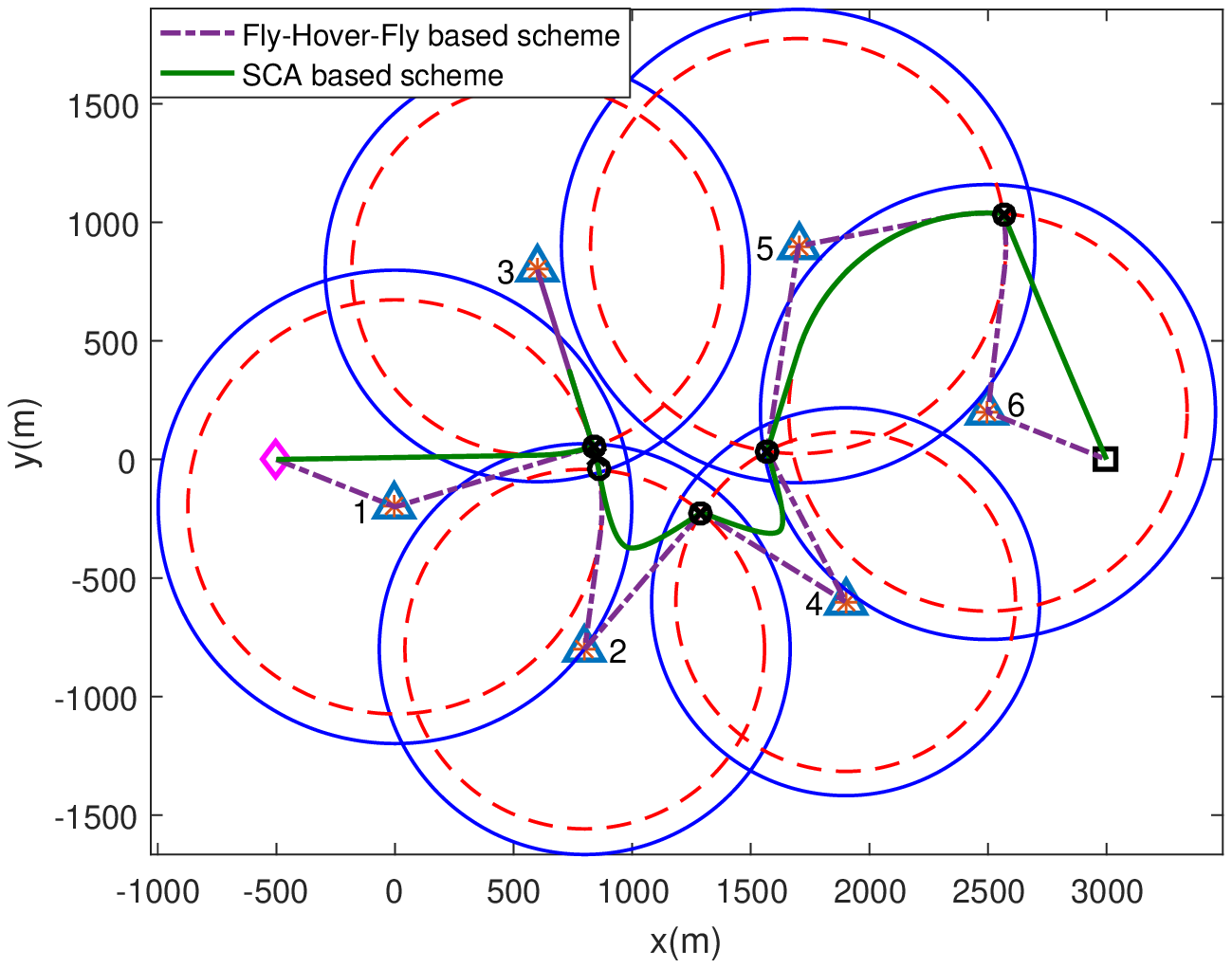}}
\subfigure[Hybrid ${\bm \theta}_1$ and ${\mathbf{U}}_1$.]{\label{QoS03hyb}
\includegraphics[width= 2.3in, height=1.8in]{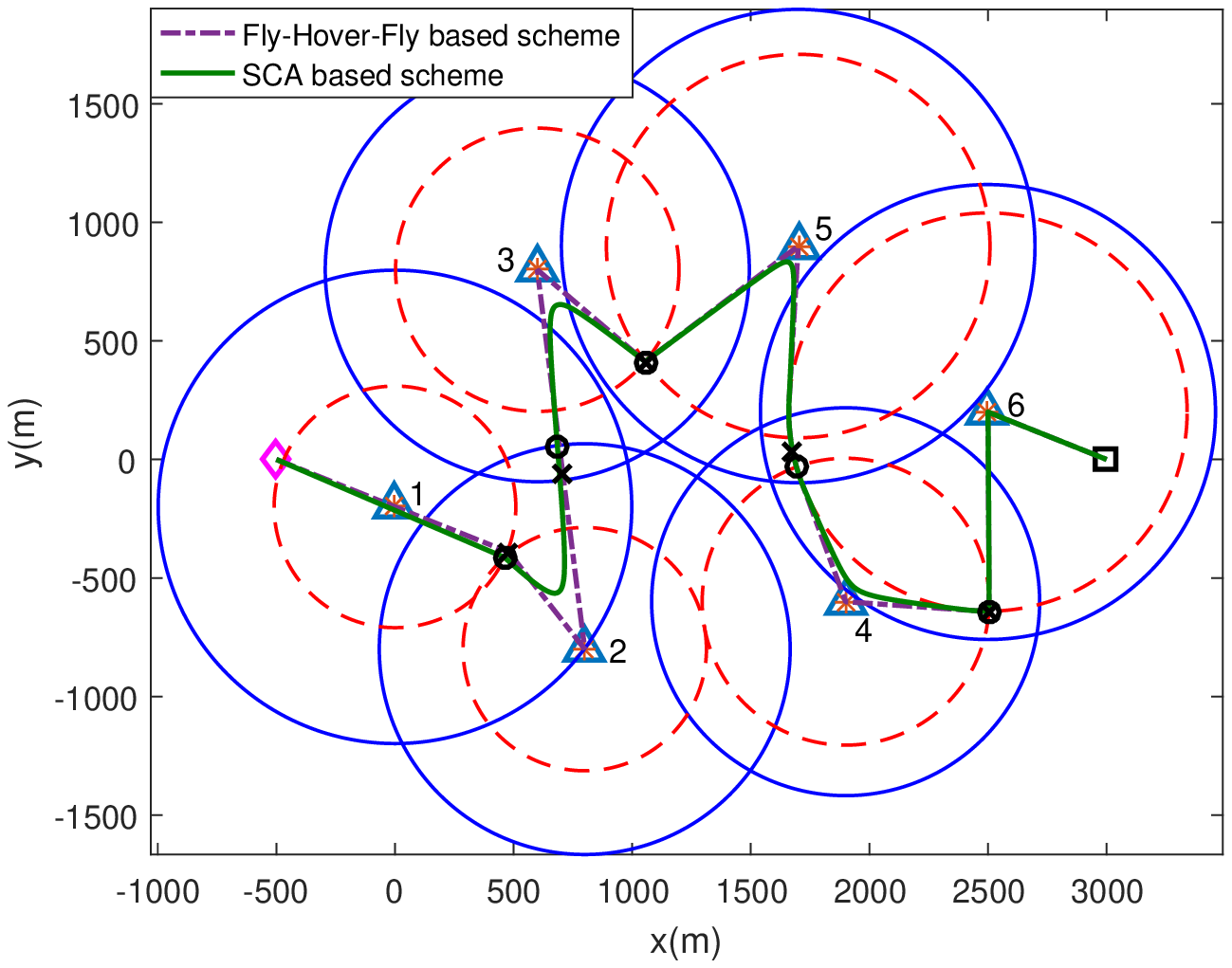}}
\quad
\subfigure[$\theta$=0.3 bit/s/Hz and $U$=80 Mbits.]{\label{QoS03Data80}
\includegraphics[width= 2.3in, height=1.8in]{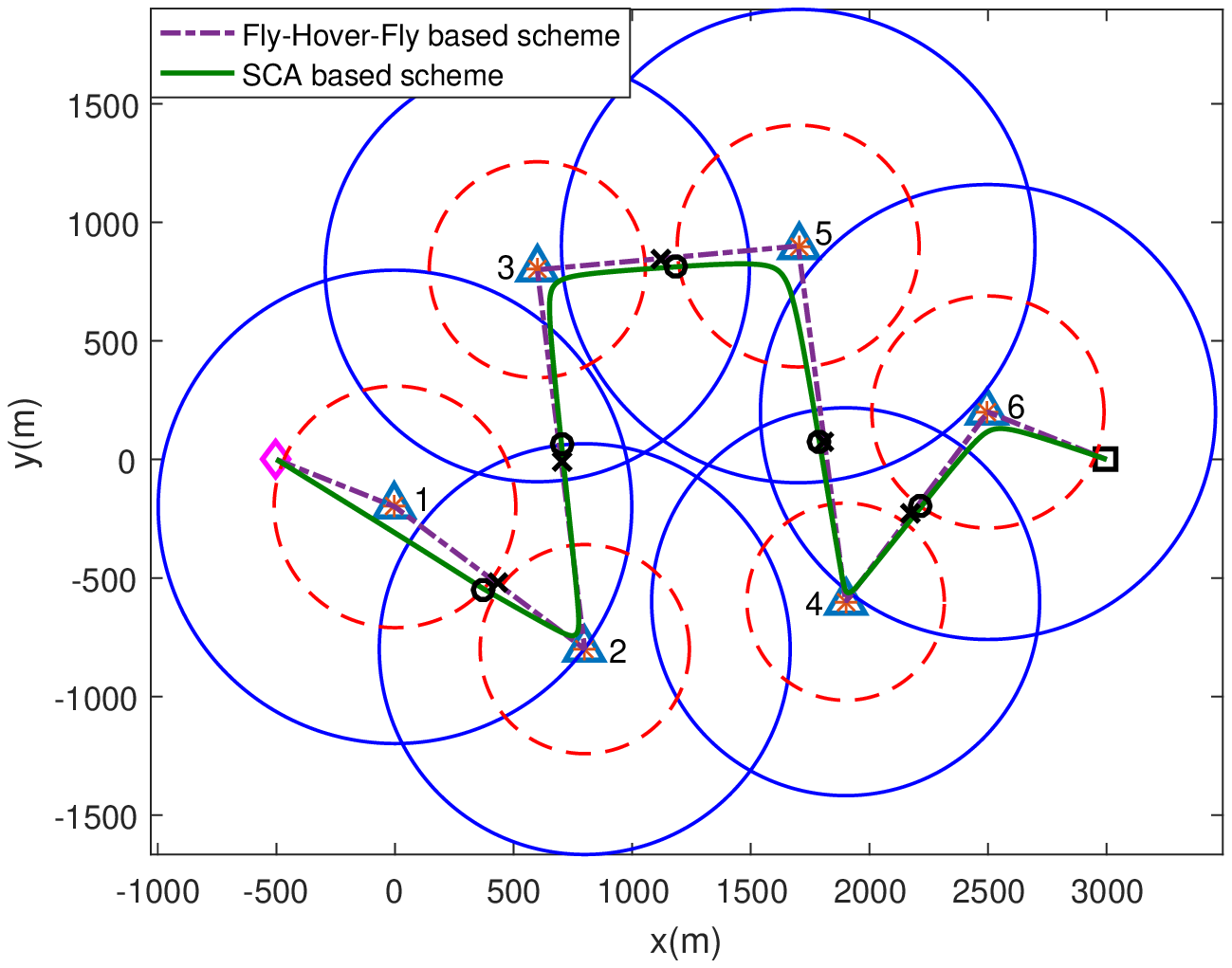}}
\subfigure[$\theta$=0.8 bit/s/Hz and $U$=80 Mbits.]{\label{QoS08Data80}
\includegraphics[width= 2.3in, height=1.8in]{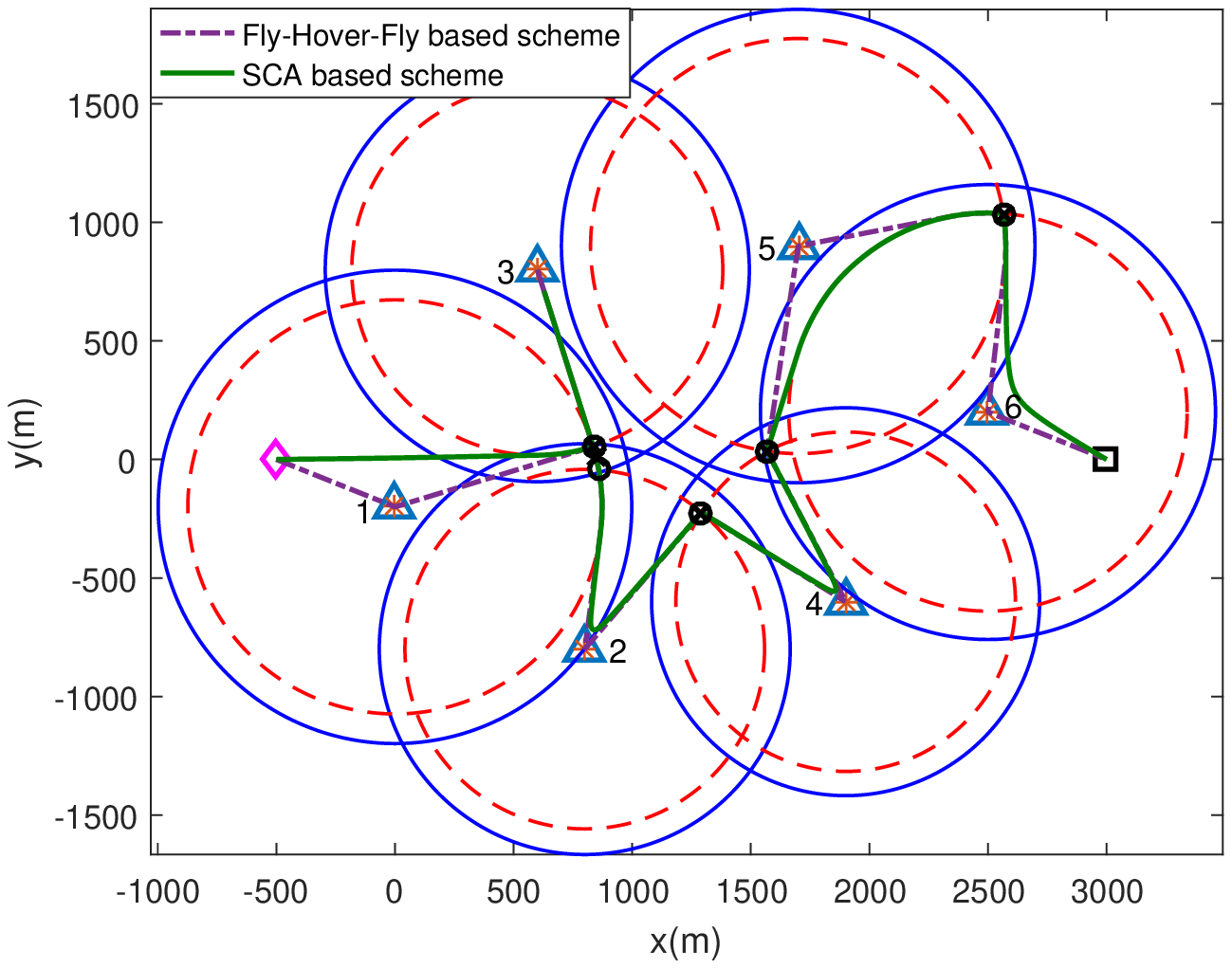}}
\subfigure[Hybrid ${\bm \theta}_2$ and ${\mathbf{U}}_2$.]{\label{QoS08hyb}
\includegraphics[width= 2.3in, height=1.8in]{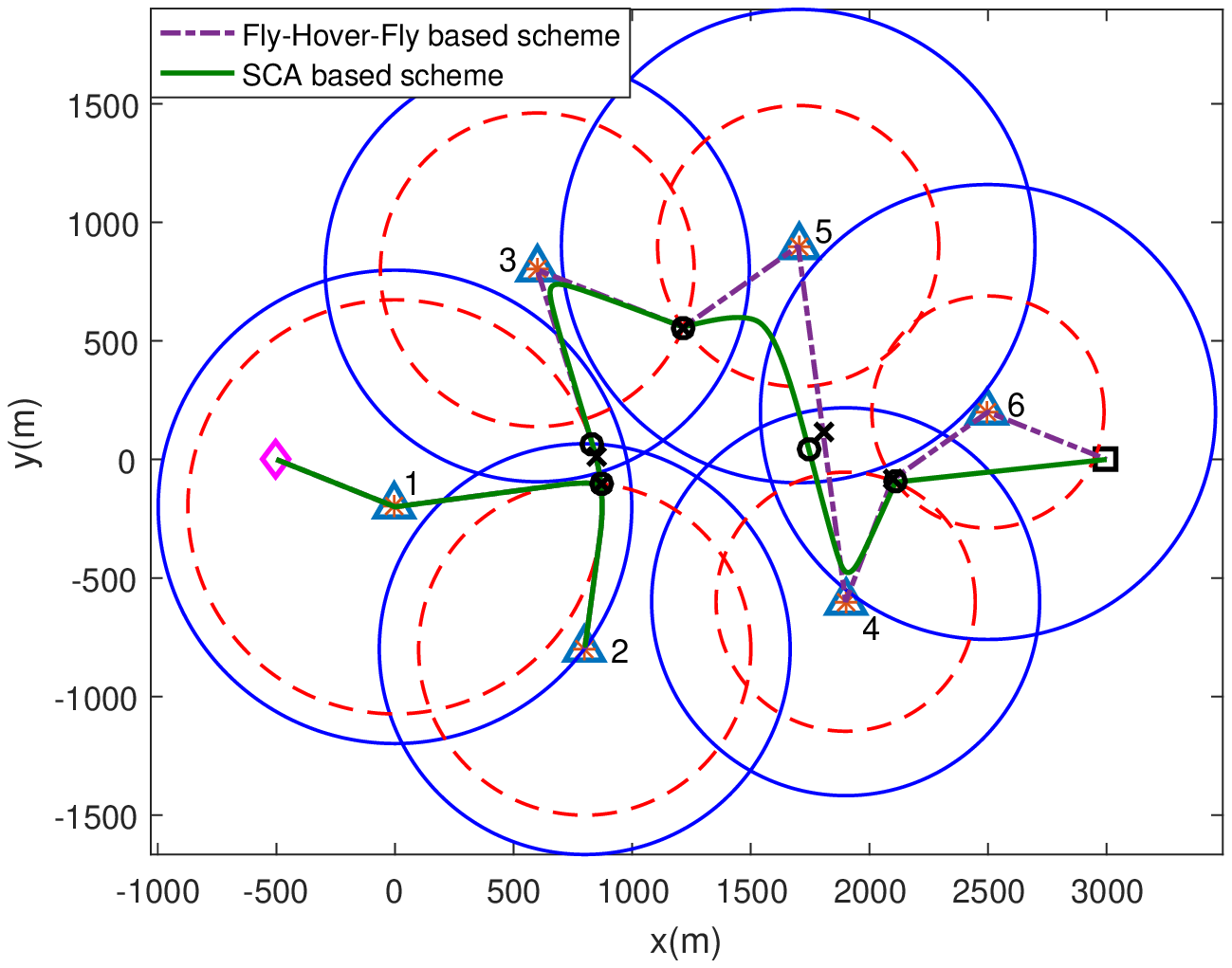}}
\setlength{\abovecaptionskip}{-0cm}
\caption{{Trajectories of the UAV with proposed designs with different configurations.}}\label{UAV trajectories1}
\end{figure*}

In this section, numerical examples are presented to validate the performance of proposed schemes. We consider that $M=6$ GBSs, with their locations shown as ``$\vartriangle $'' in Fig. \ref{UAV trajectories1}. {As described before, the terrestrial and air-to-ground channels are modeled with the UMa scenario in 3GPP technical reports~\cite{3GPP_UAV,3GPP}, the simulation parameters are set as follows: The height of each GBSs and GUEs are fixed as $H^G=25$ m and $H^{GUE}=1.5$ m. The UAV height is fixed as $H^G=110$ m, which complies with the LoS channel assumptions and the maximum allowed altitude (122m) in federal aviation authority (FAA) regulations. The transmit power of the UAV and each GUEs are set to be identical as $p^{UAV}=p^{UE}=23$ dBm. The total communication bandwidth is $W=1$ MHz. The noise spectrum density at the receiver is $- 174$ dBm/Hz. The down-tilt angle of the GBSs and the vertical antenna beamwidth are 20 degree and 30 degree. The GBS antenna main lobe gain and side lobe gain are set as $\left( {{g_m},{g_s}} \right) = \left( {10,0.5} \right)$ dB~\cite{AzariCellular}. The cell radius is 500 m. The location of each GUE is randomly distributed in its served GBS mainlobe coverage region. The UAV's maximum speed is $V_{max}=50$ m/s.} The initial and final location of the UAV are set as ${{\mathbf{q}}_I} = {\left[ { - 500,0} \right]^T}$ and ${{\mathbf{q}}_F} = {\left[ { 3000,0} \right]^T}$ as shown in  Fig. \ref{UAV trajectories1}. Except other demonstration, it is assumed that the UAV needs to upload identical amount of information bits to different GBSs, i.e., ${\widetilde U_m} = U,m \in {\mathcal{M}_{BS}}$ and each GUEs has identical minimum QoS requirement, i.e., ${\theta _m} = \theta ,m \in {\mathcal{M}_{UE}}$. The value of $N_s$ is set to be 100 and the value of $N$ is determined by the constructed shortest path.

In Fig. \ref{UAV trajectories1}, trajectories of the UAV obtained by the fly-hover-fly based design and the SCA based trajectory design are presented with different configurations. The blue circles represent the uplink NOMA zones of each GBSs and the red dash circles represent the QoS protected zones for each GUEs. The hovering locations $\left\{ {{{\mathbf{q}}_m}} \right\}$ with different GBSs are shown as ``$*$''. The UAV handover locations obtained by the fly-hover-fly based design and the SCA based trajectory design are noted as ``$\circ $'' and ``$\times  $'', respectively. Since the UAV path designed by the fly-hover-fly based scheme is only determined by the initial/final location and hovering locations with different $\left\{ {D_m^{NOMA}} \right\}$ and $\left\{ {D_m^{QoS}} \right\}$, it remains unchanged with different $U$ under same $\theta$. In contrast, the trajectory designed by the SCA based scheme is more flexible as the SCA based scheme designs the UAV trajectory considering the amount of required information bits. As shown in Fig. \ref{QoS03Data20}, when $U=20$ Mbits and $\theta =0.3$ bit/s/Hz, the UAV trajectory designed by the SCA based scheme is more likely to fly in a straight line from ${{\mathbf{q}}_I}$ to ${{\mathbf{q}}_F}$. On one hand, a small amount of required uploading information bits makes the UAV do not have to fly towards to the hovering locations $\left\{ {{{\mathbf{q}}_m}} \right\}$ which achieve the highest communication rate. On the other hand, the small QoS protected zones have little influence on the design of the UAV trajectory. As shown in Fig. \ref{QoS08Data20}, when $U=20$ Mbits and $\theta =0.8$ bit/s/Hz, the UAV needs to travel a longer distance from ${{\mathbf{q}}_I}$ to ${{\mathbf{q}}_F}$ in both schemes. In Fig. \ref{QoS03Data80} and Fig. \ref{QoS08Data80}, it is observed that the UAV trajectory obtained by the SCA based scheme are similar to that obtained by the fly-hover-fly scheme. This is expected as the UAV has a large amount of information bits to upload, the UAV needs to fly towards to the hovering locations $\left\{ {{{\mathbf{q}}_m}} \right\}$ to achieve better channel condition and minimize the mission completion time. It also indicates the optimality of the fly-hover-fly based design when $U$ increases. Furthermore, Fig. \ref{QoS03hyb} and Fig. \ref{QoS08hyb} provide the obtained UAV trajectory of our proposed designs with different values of $\widetilde{U}_m$ and $\theta_m$. Define ${\bm \theta}  = \left[ {{\theta _1},{\theta _2}, \cdots ,{\theta _M}} \right]$ and ${\mathbf{U}} = \left[ {{{\widetilde U}_1},{{\widetilde U}_2}, \cdots ,{{\widetilde U}_M}} \right]$. In Fig. \ref{QoS03hyb}, we set the parameters as ${\bm \theta}_1  = \left[ {0.3,0.4,0.5,0.6,0.7,0.8} \right]$ bit/s/Hz and ${\mathbf{U}}_1 = \left[ {20,40,60,80,100,120} \right]$ Mbits. In Fig. \ref{QoS08hyb}, ${\bm \theta}_2  = {\rm {fliplr}}\left( {\bm \theta}_1  \right)$ and ${\mathbf{U}}_2  = {\rm {fliplr}}\left( {\mathbf{U}}_1  \right)$, where ${\rm {fliplr}}\left(  {\mathbf{a}}  \right)$ denotes the manipulation of reversing the order of elements of vector ${\mathbf{a}}$. As illustrated, the UAV needs to travel a longer distance to hand over between cells which contain a higher QoS requirement GUE due to the more strict constraints. It is also observed that the UAV trajectory obtained by the SCA based scheme becomes more similar with that obtained by the fly-hover-fly based scheme as $\widetilde{U}_m$ increases.
\begin{figure}[t!]
    \begin{center}
        \includegraphics[width=3in]{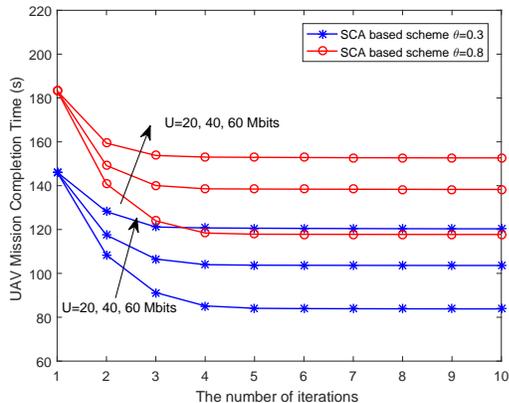}
        \caption{{Convergence of Algorithm 2 for UAV mission completion time minimization.}}
        \label{Convergence}
    \end{center}
\end{figure}
\begin{figure}[t!]
    \begin{center}
        \includegraphics[width=3in]{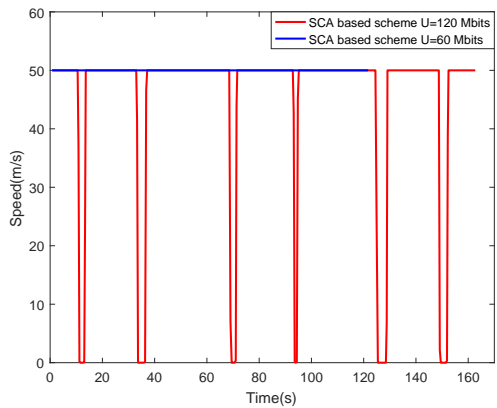}
        \caption{{UAV speed versus time with the SCA based trajectory design, $\theta =0.3$ bit/s/Hz.}}
        \label{UAV speed}
    \end{center}
\end{figure}

In Fig. \ref{Convergence}, we show the convergence of the proposed SCA based trajectory design with different parameters. The initial UAV trajectory and UAV-GBS association vectors are obtained from the fly-hover-fly based design. It is observed that the proposed SCA scheme converges quickly with only a few number of iterations. The objective value of the SCA based scheme is always lower than that of the fly-hover-fly based scheme, which is consistent with \textbf{Remark \ref{remark:SCA mode}}. In addition, when $U$ increases, the SCA based scheme converges more quickly and achieves less performance gains compared with the fly-hover-fly based scheme. It implies the asymptotically optimal feature of the fly-hover-fly based scheme.

In Fig. \ref{UAV speed}, the instant UAV flying speed of the SCA based scheme with different $U$ is presented. It is observed that the UAV flying speed is consistent with \textbf{Theorem \ref{fly-hover-fly policy}}. When $U$ is 60 Mbits, the UAV always flies at $V_{max}$ which is a special case of the proposed fly-hover-fly structure. This is because unless hovering at $\mathbf{q}_m$, the UAV can always use the hovering time to fly towards $\mathbf{q}_m$ which achieves higher communication rate and less mission completion time. When $U$ is 120 Mbits, it is expected that the UAV flies as the general proposed fly-hover-fly structure since $U$ is large and the UAV needs to hover at each $\mathbf{q}_m$ to achieve less mission completion time.

\begin{figure}[t!]
\centering
\includegraphics[width=3in]{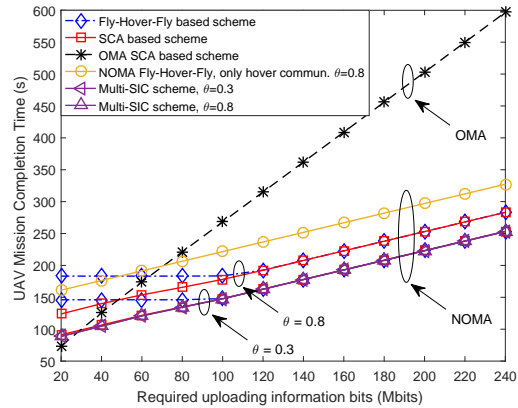}
\caption{Mission completion time versus required uploading information bits $U$.}
\label{Throughput requirement}
\end{figure}

In Fig. \ref{Throughput requirement}, the results of mission completion time versus required information bits $U$ with different schemes are presented. Besides the proposed schemes, three benchmark schemes are considered:
\begin{itemize}
  \item \textbf{Multi-SIC}: In this scheme, the SIC can also be performed at the non-associated GBSs as described in \textbf{Remark \ref{multi SIC define}}. The problem is solved with the method in \textbf{Remark \ref{remark:multi-SIC}}.
  \item \textbf{NOMA fly-hover-fly, only hovering communicate}: In this scheme, the UAV only communicates when hovering, the path is designed to find the shortest distance to visit all $\mathbf{q}_m$ from ${{\mathbf{q}}_I}$ to ${{\mathbf{q}}_F}$ without considering uplink NOMA zones and QoS protected zones. This problem can be solved as a TSP problem with predefined start and end points.
  \item \textbf{OMA SCA based trajectory design}: In this scheme the UAV and GUEs are assigned with equal bandwidth and the trajectory was designed without considering uplink NOMA zones and protected zones as did in~\cite{Zeng2019Energy}.
\end{itemize}
It is firstly observed that the performance of the SCA based scheme always outperforms the fly-hover-fly scheme especially for small amount of $U$, which is consistent with \textbf{Remark \ref{remark:SCA mode}}. However, when $U$ increases, the UAV mission completion time achieved by the fly-hover-fly based scheme and the SCA based scheme are almost the same. This is expected owing to the asymptotically optimal feature of the fly-hover-fly scheme for large $U$. Moreover, it seems that the UAV mission completion time and the amount of required uploading information bits are linearly related as $U$ increases, which is also consistent with the expression in equation \eqref{fly-hover-fly time}. The UAV mission completion time achieved by the fly-hover-fly scheme remains unchanged for small $U$. This is because the UAV mission completion time in the fly-hover-fly based design is only determined by the travelling time and the hovering time is zero for small $U$. Regarding the three benchmark schemes, it is observed that the ``Multi-SIC'' scheme achieves the same performance in both $\theta =0.3$ and 0.8 bit/s/Hz. It is expected since the ``Multi-SIC'' scheme only needs to consider the uplink NOMA zones, which remain the same for different $\theta$. Specifically, when $\theta=0.3$ bit/s/Hz, our proposed SCA based scheme is capable of achieving almost the same performance with the ``Multi-SIC'' scheme. When $\theta =$ 0.8 bit/s/Hz, the ``Multi-SIC'' scheme outperforms our proposed SCA based scheme. This is because performing the SIC at the non-associated GBSs imposes less constraints for UAV trajectory design. However, it also brings extra complexity due to additional SIC implementations at the non-associated GBSs. The results validate the effectiveness of our proposed schemes, especially for lower QoS requirements, even though the UAV interference is not canceled at the non-associated GBSs. ``NOMA fly-hover-fly, only hovering communicate'' outperforms the fly-hover-fly based scheme for small amount of $U$ when $\theta=0.8$ bit/s/Hz. This is expected as the strictly suboptimal feature of the fly-hover-fly based scheme when $U$ is not large. When $U$ increases, the performance gets worse due to the travelling time is not used for data transmission. Furthermore, it is observed that there is a significant mission completion time reduction achieved by the NOMA scheme compared with the ``OMA-SCA based trajectory design'' for large $U$. This is due to the limited bandwidth that can be used by OMA scheme. It also implies the benefit provided by NOMA in rate demanding UAV payload communication.
\begin{figure}[t!]
\centering
\includegraphics[width=3in]{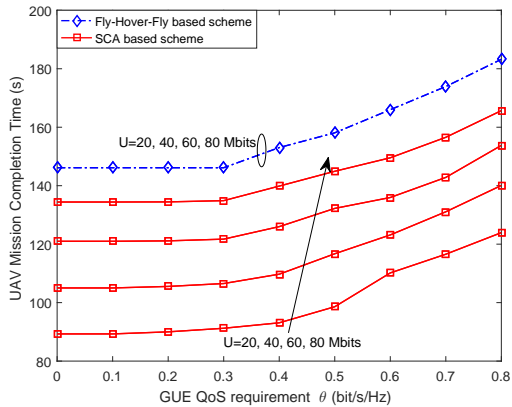}
\caption{{Mission completion time versus QoS requirement $\theta$.}}
\label{QoS requirement}
\end{figure}

Finally, as shown in Fig. \ref{QoS requirement}, the results of mission completion time versus $\theta$ under the two proposed schemes with different $U$ are presented. The UAV mission completion time of the fly-hover-fly based scheme with different $U$ are same as long as $\theta$ remains unchanged, since $U$ is not large and the mission completion time is only determined by the UAV travelling time. As expected, when $U$ increases, the UAV mission completion time achieved by the SCA based scheme increases. The UAV mission completion times of both schemes remain nearly the same for lower QoS requirement of GUEs since a smaller $\theta$ contributes smaller $\left\{ {D_m^{QoS}} \right\}$ and has little restriction on the design of the UAV trajectory. Fig. \ref{QoS requirement} further demonstrates a positive correlation between the UAV mission completion and GUEs' QoS requirements. This is because the increase of $\theta$ lead to larger $\left\{ {D_m^{QoS}} \right\}$ and thus enlarges the minimum UAV travelling distance $D_{fly}$. As a result, the UAV mission completion time increases.

\section{Conclusion}

In this paper, the UAV mission completion time minimization problem has been investigated in the ground-aerial uplink NOMA cellular networks. Specifically, the formulated problem was a mixed integer non-convex problem and non-trivial to solve directly. The feasibility of the formulated problem was efficiently checked by examining the connectivity of a carefully designed graph. Next, we first propose an efficient solution based on fly-hover-fly policy by applying graph theory techniques. Then, an iterative UAV trajectory design was proposed with SCA technique to find a high-quality suboptimal solution, which satisfies the fly-hover-fly structure. Numerical results verify the effectiveness of our proposed designs and reveal uplink NOMA transmission is an appealing solution for UAV rate demanding payload communication. Our future work may consider investigating the tradeoff between the performance and complexity of performing the SIC at non-associated GBSs. Considering the NLoS environment and online UAV trajectory design with mobile GUEs are also promising future research directions.

\section*{Appendix~A: Proof of Proposition~\ref{feasible TO P1}} \label{Appendix:A}

First, we prove the ``if'' part by showing that a feasible solution to Problem (P1) can be constructed with the UAV-GBS association order ${\bf{\Phi}} = \left[ {{\phi_1},{\phi_2}, \cdots ,{\phi_{K}}} \right]$ which satisfies \eqref{Q1}-\eqref{A}. First, with the association order, the UAV-GBS association vectors can be constructed which satisfies \eqref{One UAV Association Constraint2} and \eqref{One UAV Association Constraint2}. The UAV path can be divided into $K$ portions with $K+1$ waypoints $\left\{ {{{\mathbf{q}}_k}} \right\}_{k = 0}^{K}$, where ${{\mathbf{q}}_0} = {{\mathbf{q}}_I}$, ${{\mathbf{q}}_K} = {{\mathbf{q}}_F}$ and ${{\mathbf{q}}_k} \in {\mathcal{E}_k} \cap {\mathcal{E}_{k + 1}}$. For the $k$th portions ($k = 1,2, \cdots ,K$), ${{\mathbf{q}}_{k-1}}$ and ${{\mathbf{q}}_{k}}$ are the starting and ending waypoints, respectively. Then, we have ${{\mathbf{q}}_{k - 1}},{{\mathbf{q}}_k} \in {\mathcal{E}_{{\phi_k}}}$. Since ${\mathcal{E}_{{\phi_k}}}$ is pathwise connected, a feasible path can be always constructed in ${\mathcal{E}_{{\phi_k}}}$ with ${{\mathbf{q}}_{k-1}}$ and ${{\mathbf{q}}_{k}}$. By Allocating arbitrary feasible speed to the above constructed path, the constructed UAV trajectory and association vectors satisfy \eqref{One UAV Initial Location Constraint} and \eqref{Uplink NOMA constraint}-\eqref{One UAV Association Constraint2}. The condition \eqref{A} ensures the constraint \eqref{UAV mission} to be satisfied. As a result, the proof of the ``if'' part is completed.

Next, we prove the ``only if'' part by showing that with any feasible solution $\left\{ {{\mathbf{q}}\left( t \right),{a_m}\left( t \right),0 \le } \right.$$\left. {t \le T,m \in {{{\mathcal{M}}}_{BS}}} \right\}$ to Problem (P1), we can always construct a UAV-GBS association order ${\bf{\Phi}} = \left[ {{\phi_1},{\phi_2}, \cdots ,{\phi_{K}}} \right]$ that satisfies condition \eqref{Q1}-\eqref{A}. First, with the UAV-GBS association vectors which satisfy constraints \eqref{One UAV Association Constraint} and \eqref{One UAV Association Constraint2}, a UAV-GBS association order ${\bf{\Phi}} = \left[ {{\phi_1},{\phi_2}, \cdots ,{\phi_{K}}} \right]$ can be constructed. Constraint \eqref{UAV mission} makes the UAV associates with each GBSs at least once during the trajectory, thus condition \eqref{A} is satisfied. With the association order ${\bf{\Phi}} = \left[ {{\phi_1},{\phi_2}, \cdots ,{\phi_{K}}} \right]$, the trajectory is divided into $K$ portions with $K+1$ waypoints. Define $\left\{ {{{\mathbf{q}}_k}} \right\},k = 1,2, \cdots ,K - 1$ as the end locations of the $k$th portion trajectory and ${{\mathbf{q}}_0} = {{\mathbf{q}}_I}$, ${{\mathbf{q}}_K} = {{\mathbf{q}}_F}$. For the $k$th portion trajectory, we have ${{\mathbf{q}}_{k - 1}},{{\mathbf{q}}_k} \in {\mathcal{E}_{{\phi_k}}}$. Condition \eqref{Q1} is satisfied when $k=1$ and $k=K$. Furthermore, for each $\left\{ {{{\mathbf{q}}_k}} \right\},k = 1,2, \cdots ,K - 1$, we have ${{\mathbf{q}}_k} \in {\mathcal{E}_{{\phi_k}}}$ and ${{\mathbf{q}}_k} \in {\mathcal{E}_{{\phi_{k+1}}}}$, thus condition \eqref{QN_N+1} is satisfied. The proof of the ``only if'' part is completed.

Above all, the proof of Proposition~\ref{feasible TO P1} is completed.
\section*{Appendix~B: Proof of Theorem~\ref{fly-hover-fly policy}} \label{Appendix:B}

We prove Theorem~\ref{fly-hover-fly policy} by showing that for any feasible trajectory to Problem (P1) denoted by {${\left\{ {\mathbf{\widetilde q}\left( t \right),0 \le t \le \widetilde T} \right\}}$} that does not satisfy the proposed fly-hover-fly structure, we can always construct a new feasible trajectory to Problem (P1) which satisfies fly-hover-fly structure and achieves lower mission completion time. Particularly, denote ${\mathbf{\widetilde q}_{\max,m }} = \mathop {\max }\limits_{\mathbf{\widetilde q}\left( t \right)} \;R\left( {\mathbf{\widetilde q}\left( t \right)} \right)$ as the location along the UAV trajectory which achieves the highest communication rate when the UAV is associated with GBS $m$. Then, we can construct a new trajectory based on ${\left\{ {\mathbf{\widetilde q}\left( t \right),0 \le t \le \widetilde T} \right\}}$ by making the UAV travel at $V_{max}$. Assume that it takes $t_0$ time less to transform the original $\left\{ {\mathbf{\widetilde q}\left( t \right)} \right\}$ to the new trajectory $\left\{ {\mathbf{\widehat q}\left( t \right)} \right\}$. For the new trajectory, the remaining time $t_0$ is used to hover at ${\mathbf{\widetilde q}_{\max,m }}$ and denote the total achieved throughput as $\widehat U_m$. For the same time ${\widetilde T}$, $\widehat U_m > \widetilde U_m$, where $\widetilde U_m$ denotes the original total achieved throughput by $\left\{ {\mathbf{\widetilde q}\left( t \right)} \right\}$. In other words, the new trajectory can achieve less mission completion time than the original trajectory under the same amount of required uploading information bits.

The proof of Theorem~\ref{fly-hover-fly policy} is completed.
\section*{Appendix~C: Proof of Proposition~\ref{optimal hovering position}} \label{Appendix:D}

First, we consider Problem (P2) without constraint~\eqref{subproblem constraints2}. The optimization problem becomes a convex problem and it is easy to know the optimal solution is ${\mathbf{q}_m} = {{\mathbf{b}_m}}$. Though constraint~\eqref{subproblem constraints2} makes (P2) non-convex, ${\mathbf{q}_m} = {{\mathbf{b}_m}}$ can always achieve the minimum value as long as it is a feasible point.

Next, if ${\mathbf{q}_m} = {{\mathbf{b}_m}}$ does not satisfy constraint~\eqref{subproblem constraints2}, there exists at least one $i \in {\mathcal{M}_{BS}},i \ne m$ that ${\left\| {{{\mathbf{b}}_m} - {{\mathbf{b}}_i}} \right\|^2} < D_i^{QoS}$. In this case, we prove the proposition by showing that for any given feasible point ${\widetilde {\mathbf{q}}_m}$, which satisfy~\eqref{subproblem constraints1} and ${\left\| {{{\widetilde {\mathbf{q}}}_m} - {{\mathbf{b}}_i}} \right\|^2} > D_i^{QoS},\forall i \in {{{\mathcal{M}}}_{BS}},i \ne m$. We can always find a feasible point ${{\mathbf{q}}_m}$, which satisfies $\exists i \in {{{\mathcal{M}}}_{BS}},i \ne m,{\left\| {{{\mathbf{q}}_m} - {{\mathbf{b}}_i}} \right\|^2} = D_i^{QoS}$ and achieves smaller objective value of Problem (P2). We first construct a line segment between ${{{\mathbf{b}}_m}}$ and ${{{\widetilde {\mathbf{q}}}_m}}$. Any point in this line segment can be represented as ${\mathbf{q}} \left( \alpha  \right) = \alpha {{\mathbf{b}}_m} + \left( {1 - \alpha } \right){\widetilde {\mathbf{q}}_m},\alpha  \in \left[ {0,1} \right]$. Since ${{{\mathbf{b}}_m}}$ is located in infeasible regions and ${{{\widetilde {\mathbf{q}}}_m}}$ is located in feasible regions, we can find at least one point on the intersection of ${\mathbf{q}} \left( \alpha  \right)$ and the boundary of feasible regions. It means there always exists $0 < {\alpha ^*} < 1$ that ${\mathbf{q}} \left( {{\alpha ^*}} \right)$ is located at the boundary of feasible regions, which can be expressed as $\exists i \in {{{\mathcal{M}}}_{BS}},i \ne m,{\left\| {{\mathbf{q}} \left( {{\alpha ^*}} \right) - {{\mathbf{b}}_i}} \right\|^2} = D_i^{QoS}$. We have ${\left\| {{\mathbf{q}} \left( {{\alpha ^*}} \right) - {{\mathbf{b}}_m}} \right\|^2} = {\left\| {{\alpha ^*}{{\mathbf{b}}_m} + \left( {1 - {\alpha ^*}} \right){{\widetilde {\mathbf{q}}}_m} - {{\mathbf{b}}_m}} \right\|^2} = {\alpha ^*}{\left\| {{{\widetilde {\mathbf{q}}}_m} - {{\mathbf{b}}_m}} \right\|^2} < {\left\| {{{\widetilde {\mathbf{q}}}_m} - {{\mathbf{b}}_m}} \right\|^2}$.

The proof of Proposition~\ref{optimal hovering position} is completed.
\bibliographystyle{IEEEtran}
\bibliography{mybib}

\end{document}